\documentclass[twocolumn,pra,showpacs,final,superscriptaddress,10pt,aps,floatfix]{revtex4}
\usepackage[usenames]{color}
\usepackage{amsmath}
\usepackage{amssymb}
\usepackage{mathrsfs}
\usepackage{rotating}
\usepackage{bbold}
\usepackage{euscript}
\usepackage{graphicx}
\usepackage{graphics}
\usepackage{bm}
\newcommand{\delsq}{\bigtriangledown^2}

\newcommand{\ff}{\textstyle}

\begin{document}

\title{Breakup of H$_2^+$ by photon impact}

\author{Daniel  J. Haxton}
\affiliation{Chemical Sciences, Lawrence Berkeley National Laboratory, Berkeley CA 94720}


\begin{abstract}

Total and partial cross sections for breakup of ground rovibronic state of H$_2^+$ by photon impact are calculated
using the exact nonadiabatic nonrelativistic Hamiltonian without approximation.
The converged results span six orders of magnitude. 
The breakup cross section is divided into dissociative excitation and dissociative ionization.  The dissociative
excitation channels are divided into contributions from principal quantum numbers 1 through 4.  For dissociative ionization
the fully differential cross section is calculated using a formally exact expression.  These results are compared
with approximate expressions.  The Born-Oppenheimer expression for the dissociative ionization amplitude is shown
to be deficient near onset.  A Born-Oppenheimer approximation to the final state is shown to give accurate results
for kinetic energy sharing, the doubly differential cross section, 
between the electronic and internuclear degrees of freedom.   To accurately calculate the triply differential cross section, including the 
angular behavior, it is shown that nonadiabatic wave functions for both initial and final states are required at low electron
energies.

\end{abstract}

\maketitle

\begin{figure*}
\begin{tabular}{ccc}
\resizebox{0.3\textwidth}{!}{\includegraphics*[0in,0.5in][2.8in,2.3in]{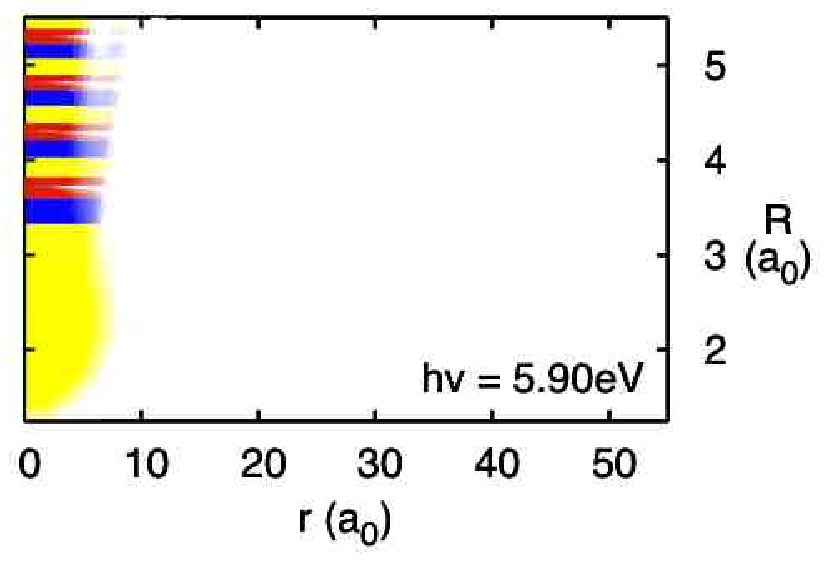}} &
\resizebox{0.3\textwidth}{!}{\includegraphics*[0in,0.5in][2.8in,2.3in]{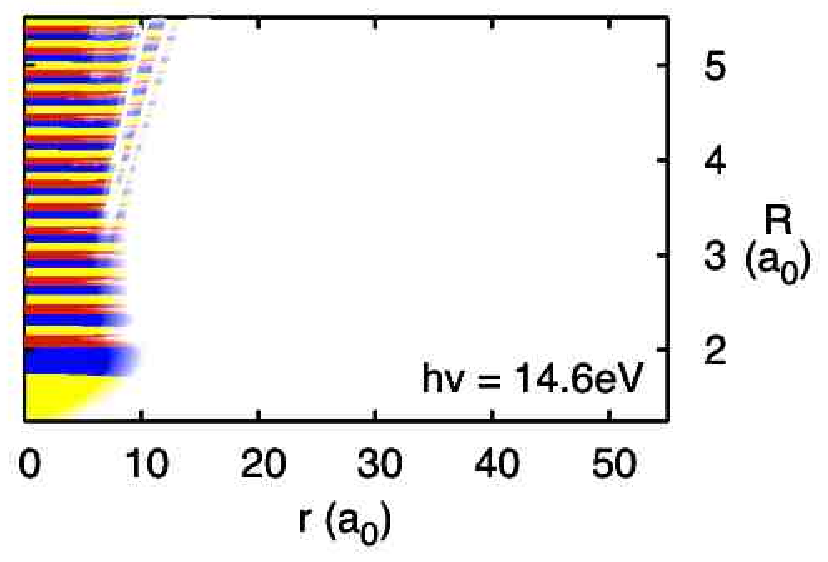}} &
\resizebox{0.36\textwidth}{!}{\includegraphics*[0in,0.5in][3.4in,2.3in]{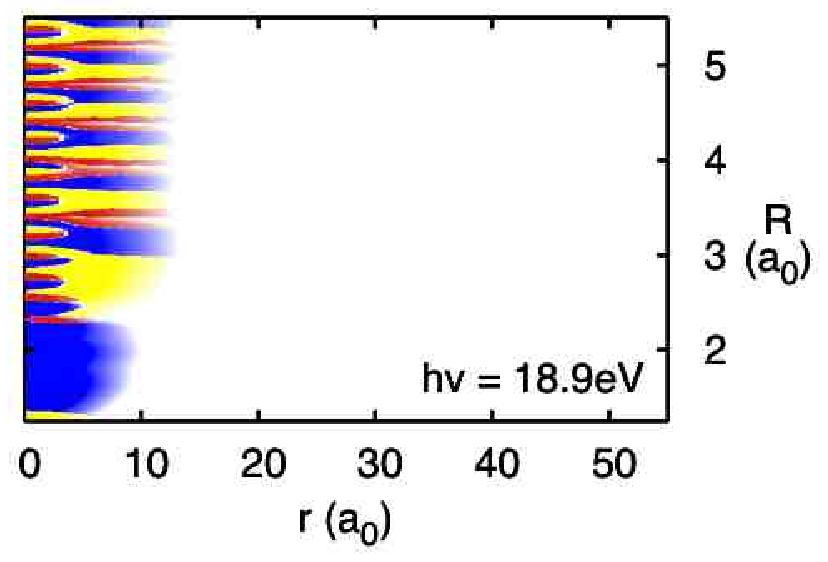}} \\
\resizebox{0.3\textwidth}{!}{\includegraphics*[0in,0in][2.8in,2.3in]{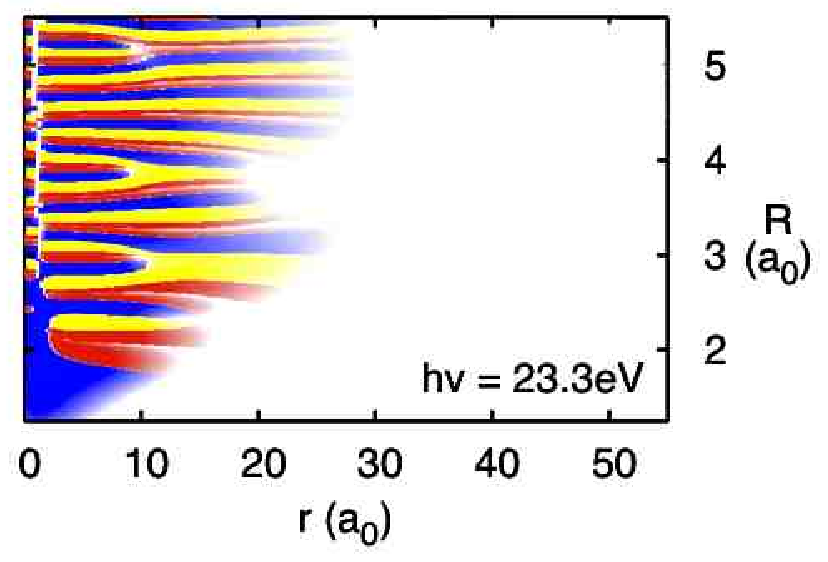}} &
\resizebox{0.3\textwidth}{!}{\includegraphics*[0in,0in][2.8in,2.3in]{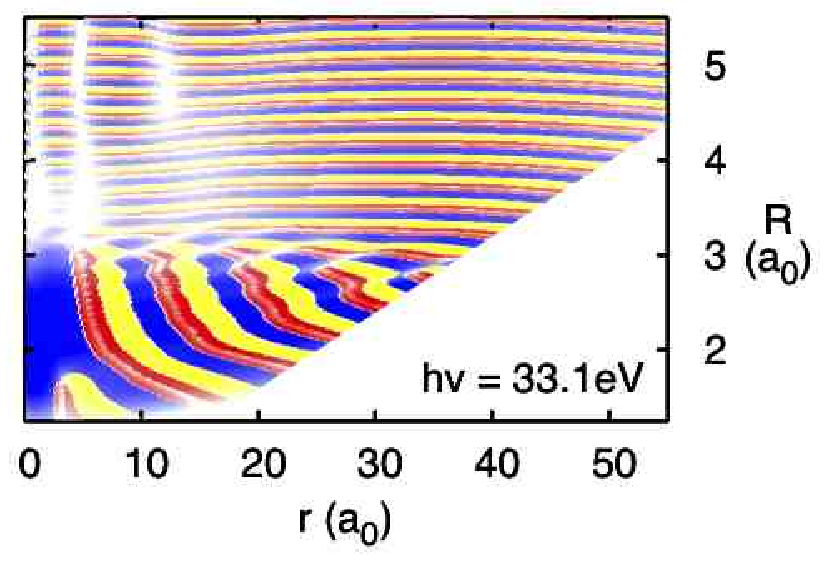}} &
\resizebox{0.36\textwidth}{!}{\includegraphics*[0in,0in][3.4in,2.3in]{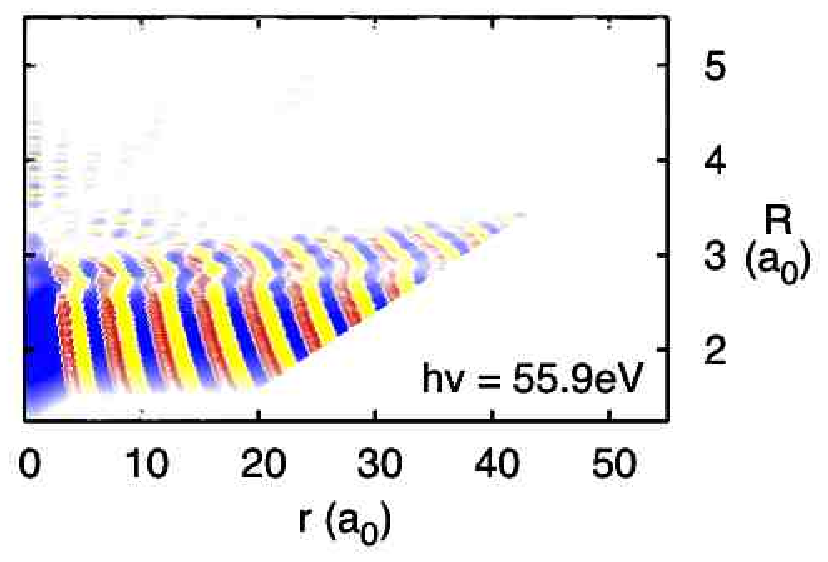}} 
\end{tabular}
\caption{(Color online) Outgoing waves $\psi^{sc}(\omega)$ for $h\nu$=5.9, 14.6, 18.95, 23.3, 33.1, and 56.0eV, evaluated at $\eta=1$, i.e., along the bond axis.  The behavior of the wave function is trivial in the $\eta$ coordinate.  The modulus is plotted on a logarithmic scale, and the color denotes the phase, with the dark-medium-light (blue-red-yellow) colors having time derivatives light-dark-medium (yellow-blue-red).  The coordinates are real valued within the plots.
\label{psiscat}}
\end{figure*}

The H$_2^+$ cation is the smallest molecule, and one that is relevant in contexts ranging from interstellar chemistry~\cite{radio,takagi}
to fusion reactors,
and as such is well studied in the literature.  
It provides the one-electron archetype for fundamental processes such as dissociative recombination~\cite{takagi}.
Due to its size, it is tractable to include nonadiabatic effects in calculating its dynamics~\cite{konoh2plus,esry1999}.
Such studies
provide insight into how coupled electronic and nuclear dynamics may be manipulated by laser light, and
in recent years the interest in
strong field and ultrafast physics has led to many experimental and numerical studies
on this topic including accurate descriptions of 
charge localization~\cite{localized,localized2},
various processes in strong fields~\cite{arbitraryh2plus, photonich2plus, chirpedh2plus,
guanh2plus,esrycomplexh2plus}, and others~\cite{highharmh2plus,ultraslowh2plus}.
Benchmark calculations of bound state rovibronic energies have been given in Refs.~\cite{beckel:3681,                                                                                                                                                                                                                                                                                                                                                                                                              
beckel:5288,QUA:QUA21735,QUA:QUA22528,QUA:QUA23070,taylorh2plus}.

The fundamental one-photon processes in H$_2^+$ may be called excitation, dissociative excitation, and dissociative ionization.  The first, 
excitation to bound vibrational states of excited Born-Oppenheimer electronic states, occurs occurs with
vanishing probability from the ground rovibrational state.  Dissociative excitation or photodissociation has been studied experimentally~\cite{buschdunn,durup,los,tadj}
and theoretically~\cite{tadj,dunn,argyros,saha}.  Dissociative ionization has received prior interest by theoreticians over the decades~\cite{poots,undulations,billtom,larkins,yabu},
and recently through the topic~\cite{cohenfano} of differential cross sections
and interference effects~\cite{geomet,fastelectrons,zeroes,pindzola,guanxray}.

Given the degree to which it is studied, it is surprising that no
\textit{ab initio} calculations of its one-photon, Fermi's golden rule breakup cross section that treat the
nuclear and electronic degrees of freedom on the same footing have been published.
The process represents one of the three fundamental Coulomb breakup problems, the others including
double ionization of helium, a complete calculation of which was reported in 2005~\cite{ecsscience}.
Here such calculations are presented.



\section{Hamiltonian and basis}

The calculations employ an implementation~\cite{prolate1_2009,prolate2_2009,prolate3_2010,haxtonmctdhf} 
of the full nonadiabatic 
Hamiltonian in prolate spheroidal coordinates~\cite{haxtonmctdhf,esry1999}.   
 The nuclear basis set is identical to that used in Ref.~\cite{haxtonmctdhf} but the expressions for the
matrix elements have been improved.
The basis functions are localized piecewise polynomials defined on a product grid in the prolate spheroidal coordinates,
with some matrix elements evaluated within the discrete variable representation (DVR) 
approximation~\cite{dickcert,coreylem,lhl,dvr00}.  For odd $\mathcal{M}$
values the basis functions include (unitless) factors of $\rho = \sqrt{(\xi^2-1)(1-\eta^2)}$ to enforce square root boundary conditions.

The exact nonrelativistic Hamiltonian may be written~\cite{johnzhang}
\begin{equation}
\begin{split}
\hat{H} = & -\frac{1}{2\mu_eR^2} \delsq + \frac{1}{R} - \frac{1}{r_A} - \frac{1}{r_B} + \frac{1}{\mu_R} \ \times \\
&  \left[ \hat{T_R} +  \frac{ J(J+1) - 2J_z^2 + \hat{J^+}\hat{l^-} + \hat{J^-}\hat{l^+} + \hat{l}^2}{2R^2} \right] \quad ,\\
& \mu_e=\frac{2\times1836.152701}{2\times1836.152701+1} \quad
\mu_R = \frac{1}{2} 1836.152701
\quad , 
\end{split}
\end{equation}
with the interparticle distances $R$, $r_A$, and $r_B$, with $\delsq$ the Laplacian in the electronic coordinates, 
and $J_z$ the projection of angular momentum (total, $J$, and electronic, $l$; the projection of nuclear angular momentum is zero) upon the bond axis, conjugate to the third Euler angle $\gamma$.  Except for $\delsq$ operators are denoted with hats and scalars have no hats in this equation.
%
For $R^{5/2}$ times the wave function the nuclear kinetic energy may be written
\begin{equation}
\begin{split}
\hat{T_R} & \ff =  -\frac{1}{2}\frac{\partial^2}{\partial R^2} + \left(\frac{1}{R} \frac{\partial}{\partial R} - \frac{1}{2R^2}\right) \left( \hat{Y} + \frac{3}{2} \right) - \frac{1}{2R^2}\left( \hat{Y} + \frac{3}{2} \right)^2 \quad ,
\end{split}
\label{tsubr}
\end{equation}
in which expression
\begin{equation}
\hat{Y} = \ff \frac{1}{\xi^2 - \eta^2} \left(\xi(\xi^2-1) \frac{\partial}{\partial \xi} + \eta(1-\eta^2) \frac{\partial}{\partial \eta} \right) \quad .
\end{equation}
It may be shown that
\begin{equation}
\begin{split}
\ff (\xi^2-\eta^2) & \ff \left[(\hat{Y}+3/2) (\hat{Y}+3/2) + l^2\right]  = \frac{9}{4}(\xi^2-\eta^2)  \\
& \ff +
%
%
 \frac{\partial}{\partial \xi}  (\xi^4 - \xi^2)  \frac{\partial}{\partial \xi} 
%
%
 +\frac{\partial}{\partial \eta} (\eta^2 -\eta^4) \frac{\partial}{\partial \eta}   \\
 & \ff + J_z^2 (\xi^2-\eta^2) \left( 1 + \frac{(\xi\eta +\alpha)^2}{(\xi^2-1)(1-\eta^2)}\right) \quad ,
\end{split}
\label{ysqexpand}
\end{equation}
in which the second derivatives $\frac{\partial^2}{\partial \xi \partial \eta}$ cancel each other, and more trivially that
\begin{equation}
\begin{split}
& \ff (\xi^2-\eta^2) \left( \hat{Y}+\frac{3}{2} \right) = \\
& \ff \left[ \xi(\xi^2-1) \frac{\partial}{\partial \xi} - \frac{1}{2} + \frac{3\xi^2}{2} \right] 
 -\left[\eta(\eta^2-1) \frac{\partial}{\partial \eta} - \frac{1}{2} + \frac{3\eta^2}{2} \right] \quad .
\end{split}
\label{yexpand}
\end{equation}

The raising and lowering operators are
\begin{equation}
\begin{split}
\ff l^\pm & \ff = \pm e^{\pm i \phi} \frac{\rho}{\xi^2-\eta^2} \left(\eta\frac{\partial}{\partial \xi} - \xi \frac{\partial}{\partial \eta}\right) +  
i \frac{\xi\eta}{\rho} \frac{\partial}{\partial \phi} \quad . \\
\end{split}
\end{equation}
As mentioned above, the primitive basis functions are defined with factors of $\rho = \sqrt{(\xi^2-1)(1-\eta^2)}$ for odd $\mathcal{M}$.  
The matrix elements of the raising and lowering operators for a bra-ket with the ket having even quantum number $\mathcal{M}$
therefore 
involve integrals of the following operator
with respect to the polynomial basis functions
\begin{equation}
\begin{split}
\ff \rho(\xi^2-\eta^2) &  l^\pm
 =\ff   
 \pm \    \bigg((1-\eta^2) \eta \left[ (\xi^2-1)\frac{\partial}{\partial \xi} + \xi\right]  \\
 & \ff - (\xi^2-1) \xi \left[ (\eta^2-1)\frac{\partial}{\partial \eta} + \eta\right]  \bigg) 
 - (\mathcal{M}\pm1) \xi \eta \quad .
\end{split}
\end{equation}
The matrix elements for 
odd-$\mathcal{M}$ ket  are in turn expressed in terms of matrix elements of the operator
\begin{equation}
\begin{split}
\ff (\xi^2-\eta^2) &  l^\pm \rho
 =\ff  
 \pm \    \bigg(... \bigg) 
 - \mathcal{M} \xi \eta \quad .
 \label{pmeqn}
\end{split}
\end{equation}

The matrix elements of the individual terms in Eq.(\ref{ysqexpand}), which are hermitian, of the operator in Eq.(\ref{yexpand}) and 
Eq.(\ref{pmeqn})
 that occurs
both in $\eta$ and $\xi$, which is antihermitian, and of the antihermitian 
$(\frac{1}{R}\frac{\partial}{\partial R} - \frac{1}{2R^2})$ operator are integrated exactly by quadrature.
As in Refs.~\cite{prolate1_2009,haxtonmctdhf} only one dimensional integrals are involved and therefore the
matrix representations of these operators are quite sparse.

The basis set employs exterior complex scaling~\cite{abc1,abc2,moi2,moi3,moi4,moirev,topicalreview04} 
of the electronic and nuclear coordinates in order to enforce outgoing wave
boundary conditions exactly.  The coordinates of electrons and nuclei are rotated into the complex coordinate plane in the asymptotic
region, which results in an antihermitian component of the Hamiltonian that only absorbs outgoing flux.  Bound states are not absorbed
and despite the fact that Rydberg states penetrate into the complex scaled region, their analytic continuations are accurately represented, 
obeying an orthogonality relationship, having unperturbed real eigenvalues, etc.

\section{Total cross sections}

The absorption cross section is calculated~\cite{accurate, practical,topicalreview04} 
by solving the linear equation
\begin{equation}
\psi^{sc}(\omega) = \hat{G}^+(E_0+\omega) \mu \psi_0 \quad ,
\label{outgoing}
\end{equation}
in which $E_0$ is the energy of the initial ground rovibronic eigenstate $\psi_0$, $\omega$ is the photon energy, $\mu$ is the dipole operator,
 and $\hat{G}^+(E_0+\omega)$ is the outgoing wave Green's function as represented
by exterior complex scaling.  

Examples of the calculated time independent half-scattering wave functions $\psi^{sc}$ 
are shown in Fig.~\ref{psiscat}.  The wave functions are evaluated parallel or perpendicular to
the bond axis for the corresponding laser polarizations, i.e., at the point $\eta=\pm 1$ or 0, correspondingly.  At the energies studied,
the behavior in $\eta$ 
is mostly uninteresting, being mostly $p$-wave outgoing flux for dissociative ionization, for instance.  

To extract the cross sections from the $\psi^{sc}$ the method of Ref.~\cite{jae96:6778} as adapted to exterior complex scaling in Ref.~\cite{haxtonmctdhf} 
is applied.
As the outgoing wave at a given photon energy is directly calculated via Eq.~\ref{outgoing}, no Fourier transform
is needed.  The total breakup cross section is obtained in the length gauge and for polarization parallel to the bond axis via
\begin{equation}
\sigma(\omega) = \frac{8}{3}\pi \alpha\omega \hbar \left\langle \psi^{sc}(\omega) \vert \mathrm{a}(\hat{H}) \vert \psi^{sc}(\omega) \right \rangle \quad ,
\label{aaa}
\end{equation}
with $\alpha$ the fine structure constant.  
 In this expression $\mathrm{a}(\hat{H})\equiv\frac{1}{2}(\hat{H}-\hat{H}^\dag)$ is the antihermitian part of 
the Hamiltonian, the hermitian part being $\mathrm{h}(\hat{H})\equiv\frac{1}{2}(\hat{H}+\hat{H}^\dag)$.
This is an isotropic cross section so there is the factor of $\frac{1}{3}$.  For perpendicular
polarization there is a factor of $\frac{2}{3}$ and so the corresponding coefficient in Eq.(\ref{aaa}) is $\frac{16}{3}$.

To distinguish dissociative excitation from dissociative ionization the antihermitian part of the hamiltonian is divided into the part
that absorbs flux for large bond lengths $R$ and that which does so for large values of the prolate spheroidal coordinate $\xi$.
With the identity
\begin{equation}
\frac{1}{r_A} + \frac{1}{r_B} = \frac{4\xi}{R(\xi^2-\eta^2)} 
\end{equation}
and the shorthand
\begin{equation}
\begin{split}
 B = & -\frac{1}{2\mu_e} \delsq + \frac{1}{2\mu_R} \left[ \left(Y+\frac{3}{2}\right)^2 + \hat{l}^2\right] \\
V = & -\frac{4\xi}{\xi^2-\eta^2} \qquad  \qquad  \ 
 D = \frac{1}{\mu_R}\left(\frac{1}{R}\frac{\partial}{\partial R} - \frac{1}{2R^2} \right)
\\
 T = & -\frac{1}{2}\frac{\partial^2}{\partial R^2} + \frac{1}{R} +\frac{J^2-2J_z}{2R^2}\qquad 
Y =  \left( \hat{Y} + \frac{3}{2}\right)  
\end{split}
\label{abbreqn}
\end{equation}
the full Hamiltonian may be abbreviated
\begin{equation}
H = \frac{1}{R^2}B + T + \frac{1}{R}V + DY 
\label{abbr}
\end{equation}
and the antihermitian part of the Hamiltonian divided
\begin{equation}
\mathrm{a}(H) = H_{anti}^e + H_{anti}^R
\end{equation}
such that
\begin{equation}
\begin{split}
\ff H_{anti}^e & \ff = h\left(\frac{1}{R^2}\right)a(B) + h\left(\frac{1}{R}\right)a(V) + a(D)h(Y) \\
\ff H_{anti}^R & \ff = a\left(\frac{1}{R^2}\right)h(B) + a(T) + a\left(\frac{1}{R}\right)h(V) + h(D)a(Y) \quad .
\end{split}
\label{shorthand}
\end{equation}
%
%
%
The  cross sections for dissociative ionization $\sigma^{DI}(E)$ and dissociative excitation $\sigma^{DE}(E)$ 
are  calculated as 
\begin{equation}
\begin{split}
\sigma^{DI}(\omega) & = \frac{8}{3}\pi \alpha\omega \hbar \left\langle \psi^{sc}(\omega) \vert H_{anti}^e \vert \psi^{sc}(\omega) \right \rangle \\
\sigma^{DE}(\omega) & = \frac{8}{3}\pi \alpha\omega \hbar \left\langle \psi^{sc}(\omega) \vert H_{anti}^R \vert \psi^{sc}(\omega) \right \rangle 
\end{split}
\label{spliteq}
\end{equation}
with $\sigma(\omega)=\sigma^{DI}(\omega)+\sigma^{DE}(\omega)$.  

\begin{figure}
\begin{tabular}{c}
\resizebox{0.7\columnwidth}{!}{\includegraphics*[1in,1.2in][5.6in,4in]{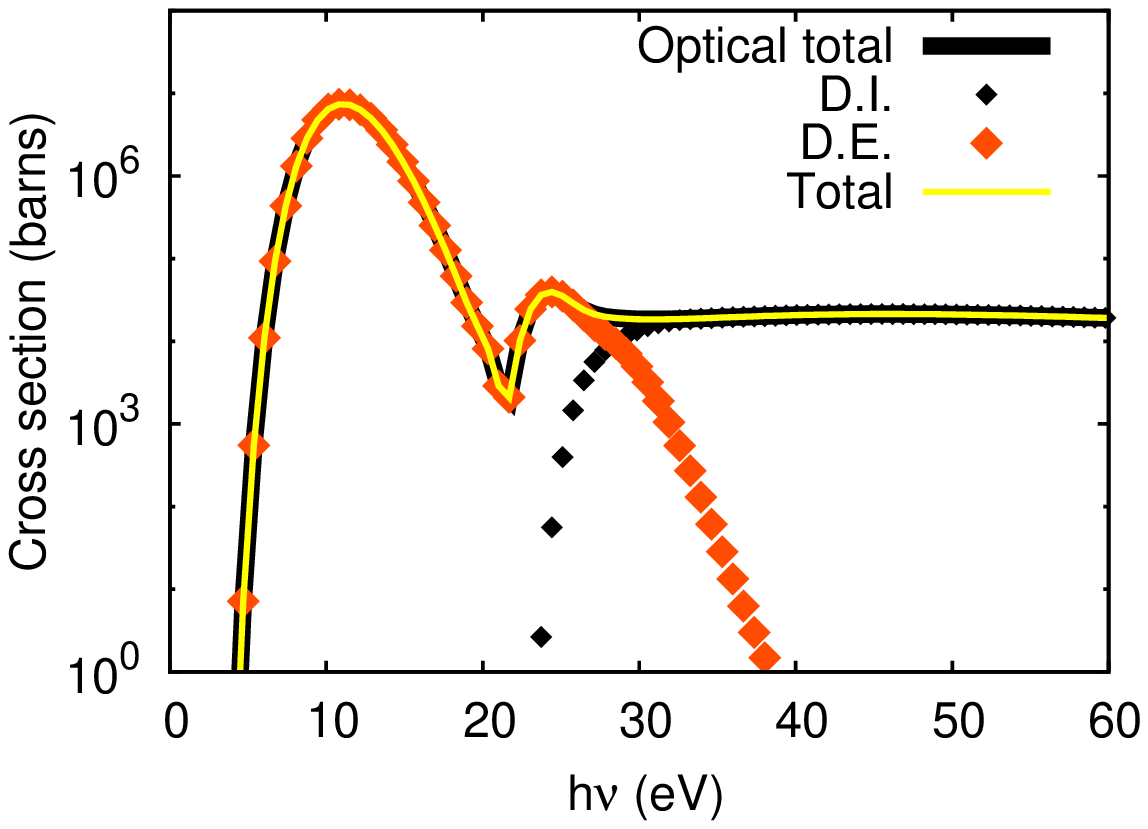}} \\
\resizebox{0.7\columnwidth}{!}{\includegraphics*[1in,0.7in][5.6in,4in]{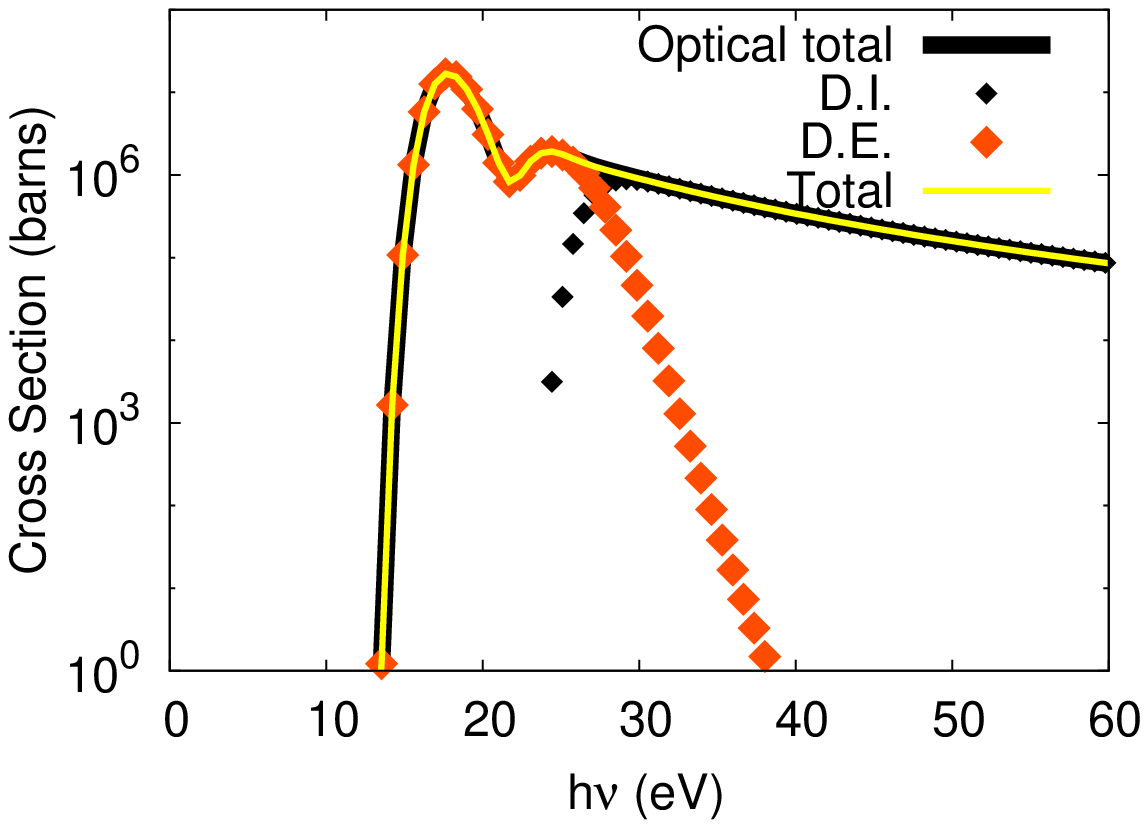}}
\end{tabular}
\caption{(Color online) Cross sections calculated for total  $\sigma(\omega)$, light colored lines, dissociative ionization $\sigma^{DI}(\omega)$,
black small diamonds, 
and dissociative excitation $\sigma^{DE}(\omega)$, large orange (pale) diamonds, as calculated via Eqs.~(\ref{aaa},\ref{spliteq}), and
total via the optical theorem, Eq.(\ref{optical}), thick black lines, for parallel (top) and perpendicular (bottom) polarization.  \label{total}}
\end{figure}

In Fig.~\ref{total} the cross sections $\sigma(E)$, $\sigma^{DI}(E)$, and $\sigma^{DE}(E)$
are plotted.
The total cross section is also calculated via the optical theorem, i.e.
\begin{equation}
\sigma(\omega)= - \frac{4}{3}\pi\alpha\omega \hbar \ \mathrm{im}\left(\left\langle \mu \psi_0 \left\vert \hat{G}^+(E_0+\omega) \right\vert \mu \psi_0 \right\rangle \right) \quad .
\label{optical}
\end{equation}
The agreement between the two formally equivalent results is essentially exact (to approximately 5-8 significant figures in general) although they are calculated quite differently.

It is true that there may be outgoing flux that is absorbed in the region in which both $H_{anti}^R$ and $H_{anti}^e$ are nonzero, calling into question
the separation described above.  However, the results for $\sigma^{DI}$ and $\sigma^{DE}$ presented here are converged with respect to the complex 
scaling radii in both the electronic and nuclear coordinate.


\section{Dissociative excitation}

The dissociative excitation cross section $\sigma^{DE}$ is divided into contributions of the final electronic states of the hydrogen atom.
The wave function is projected upon the fixed-nuclei electronic eigenfunctions $\phi_i$,
\begin{equation}
\begin{split}
\psi^{sc}_i(\xi,\eta,R;\omega) = & \phi_i(\xi,\eta;R) \int (\xi'^2-\eta'^2) d\xi' d\eta'  \ \times \\
&  \phi_i(\xi',\eta';R) \psi^{sc}(\xi',\eta',R;\omega) \quad .
\end{split}
\end{equation}
The division of the cross section proceeds via
\begin{equation}
\begin{split}
%
\sigma_{ij}^{DE}(\omega) = \frac{8}{3} \pi \alpha \omega \hbar & \langle \psi^{sc}_i(\omega) \vert  H_{anti}^R \vert   \psi^{sc}_j(\omega)  \rangle \quad ,
\end{split}
\label{nucsplit}
\end{equation}
such that $\sum_{ij} \sigma^{DE}_{ij}(\omega) = \sigma^{DE}(\omega)$.

If the final states $\phi_j$ were exact representations of the asymptotic states, and in the limit of large projection radius,
$\sigma^{DE}_{ii}$ of Eq.~\ref{nucsplit} would
be the formally exact cross section for Rydberg state $i$; the off-diagonal results $\sigma_{ij}$, $i \ne j$, would go to zero.

\subsection{Formal and numerical considerations}

However, because the prolate spheroidal coordinate $R$ is not exactly the same as the dissociative coordinate, due to the mass 
of the electron, the states $\phi_i(R)$ that are used for the projection are not exactly the asymptotic states; the asymptotic states 
are delocalized in $R$.  Due to the resulting nonadiabatic
coupling between the approximate states $\phi_i(R)$, nonzero
off-diagonal results $\sigma_{ij}$, $i\ne j$, are expected.  

Nonzero off-diagonal contributions are also expected if the projection is
not performed at a sufficiently large bond radius $R$, such that the different principal quantum number manifolds are still significantly
mixed by the interaction with the bare proton.  In any case, the off-diagonal ``cross-sections'' $\sigma_{ij}$, $i\ne j$, are spurious and 
should be significantly smaller than the physical cross sections $\sigma_{ii}$ for the latter to be regarded as reliable.

\begin{figure}
\begin{tabular}{c}
\resizebox{0.7\columnwidth}{!}{\includegraphics*[1in,1.2in][5.6in,4.1in]{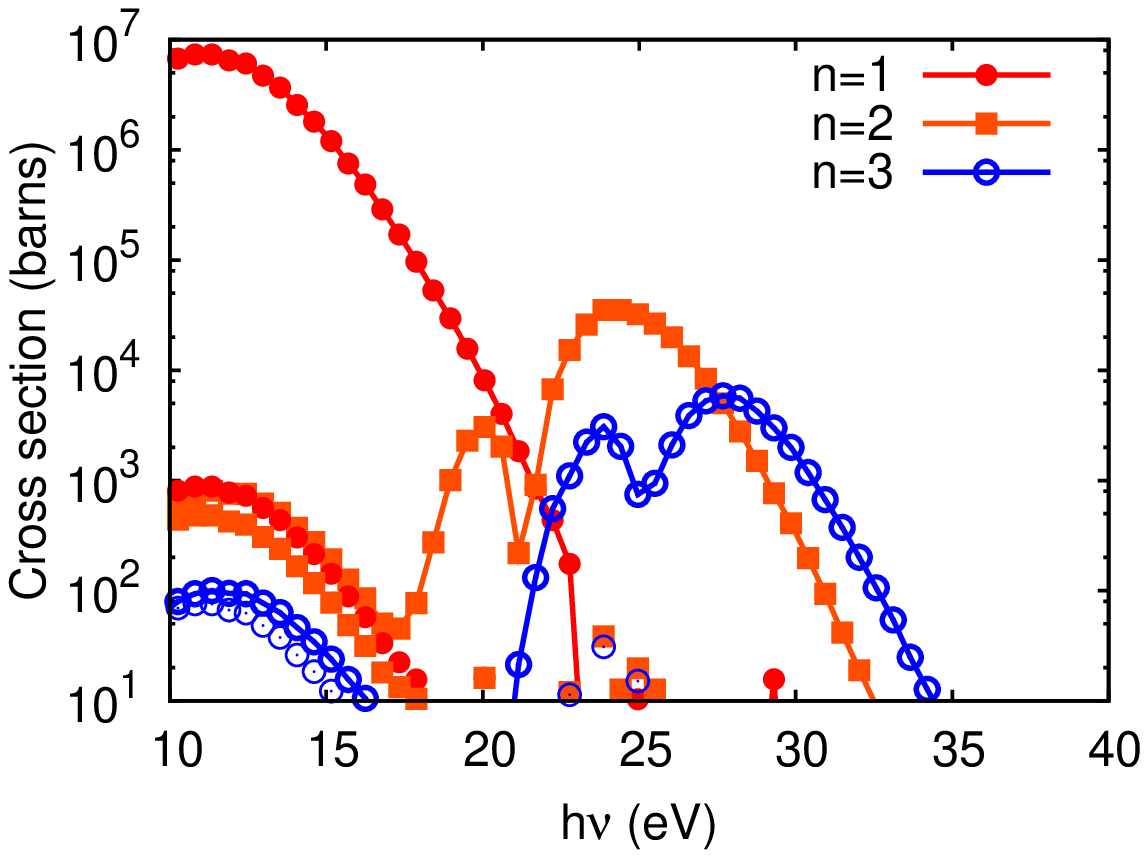}} \\
\resizebox{0.7\columnwidth}{!}{\includegraphics*[1in,0.5in][5.6in,4.1in]{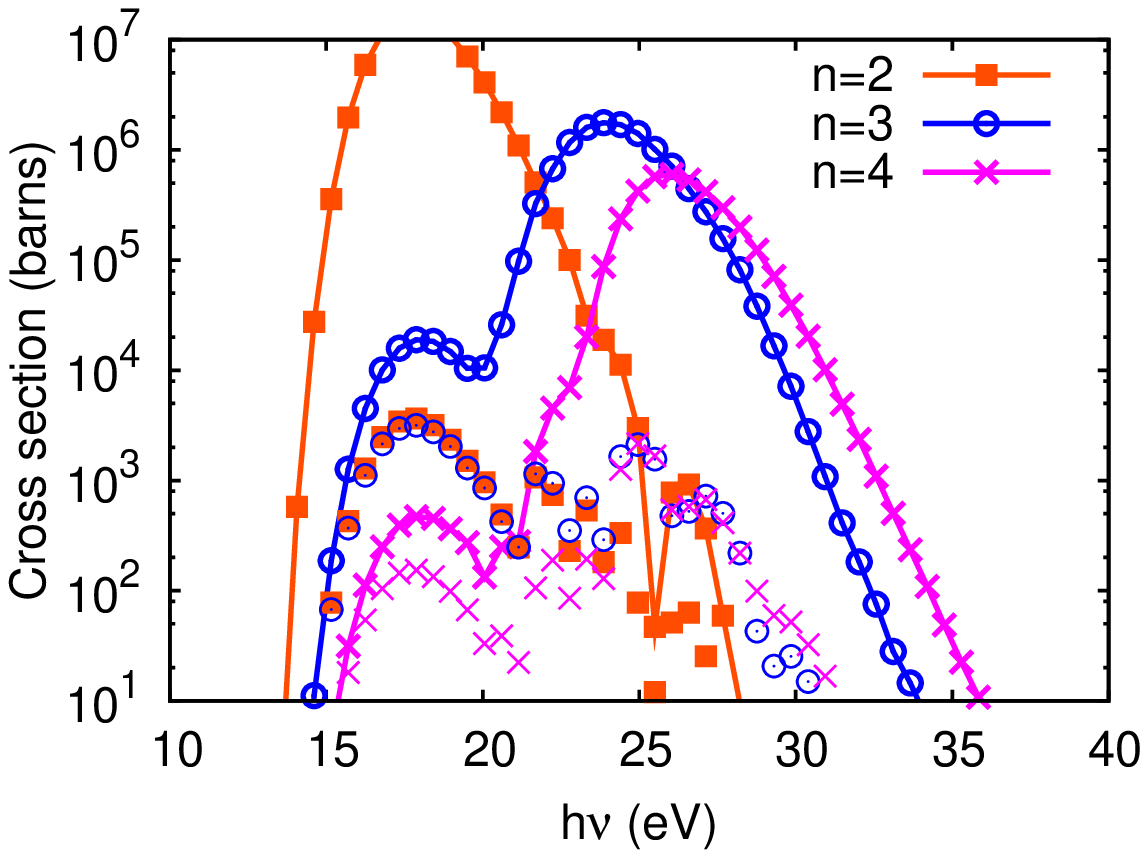}} \\
\end{tabular}
\caption{(Color online)  Cross sections $\sigma_n$ (Eq.(\ref{summ1}), connected dots) for dissociative excitation into the manifold of Rydberg states with principal
quantum number $n$, and sums of unphysical, erroneous off diagonal cross sections $\tilde{\sigma}_n$ (Eq.(\ref{summ2}), dots) as described in text.  
Top, parallel polarization; bottom, 
perpendicular.\label{nucfluxes}}
\end{figure}

The projection onto final Rydberg states must be performed at a bond length large enough such that there are no significant nonadiabatic couplings from that bond length outward.  
%
The last avoided crossing involving a $\Sigma_u$ $n=3$ electronic state
occurs at approximately 16.5$a_0$, and that involving a $\Pi_u$ $n=4$ states occurs at approximately 15$a_0$.
The calculations are observed to be well converged with an exterior complex scaling bond length slightly larger 
than these values.  The last avoided crossing involving the $n=4$ 
$\Sigma_u$ states occurs at approximately 40$a_0$ and projection upon these states was not attempted.

\subsection{Results}


Results are shown in Fig.~\ref{nucfluxes}.  This figure shows the dissociative excitation cross section binned by 
principal quantum number of the final Rydberg state.  The electronic orbital angular momentum of the final Rydberg state is not
resolved.  The total cross section into principal quantum number $n$, $\sigma^{DE}_n$ with
one subscript, is defined as
\begin{equation}
\sigma^{DE}_n = \sum_{i \in n} \sum_{j \in n} \sigma^{DE}_{ij} \quad ,
\label{summ1}
\end{equation}
with the sums of spurious cross sections off-diagonal in the principal quantum number denoted
\begin{equation}
\tilde{\sigma}^{DE}_n = \sum_{i \in n} \sum_{j \notin n} \sigma^{DE}_{ij} \quad ,
\label{summ2}
\end{equation}
in which the notation ``$\sum_{i\in n}$'' means sum over fixed-nuclei states $i$ that correlate with a given principal quantum number $n$.  As discussed, the off-diagonal sum $\tilde{\sigma}^{DE}_n$ should be regarded as a minimum error bound to the calculated physical cross section $\sigma^{DE}_n$.

As can be seen in Fig.~\ref{nucfluxes}, the largest part of the dissociative excitation cross section is the 
low energy part due to absorption into the lowest 1 $\Sigma_u$ and 1 $\Pi_u$ states.
About four orders of magnitude below the peak cross sections, at about 10 and 18eV, respectively,
one may see that there are nonzero cross sections, both diagonal and 
off-diagonal, calculated for the higher electronic states. These are nonzero below their thresholds and are congruent to the dominant 1 $\Sigma_u$ or 1 $\Pi_u$
peak.  These facts suggest that these calculated features are spurious, and probably come from contamination of the higher states' results from
the outgoing flux in the lowest electronic state channel.  As explained above, the electronic states used for the final state projection in dissociative
excitation are not precisely the long-range states, and therefore this behavior is not surprising.

%
%

At higher energy, the calculated physical cross sections $\sigma_{n}^{DE}$ in Fig.~\ref{nucfluxes} are  several
orders of magnitude above any 
unphysical off-diagonal results and therefore should be regarded as reliable.  
Nonadiabatic coupling leads to double peaks, which are especially prominent
in parallel polarization; the main peak for each final principal quantum number has a small side peak at lower 
energy due to coupling from the high energy side of the peak of the prior principal quantum number.

\begin{figure}
\begin{tabular}{c}
\resizebox{0.7\columnwidth}{!}{\includegraphics*[1in,0.5in][5.6in,4.1in]{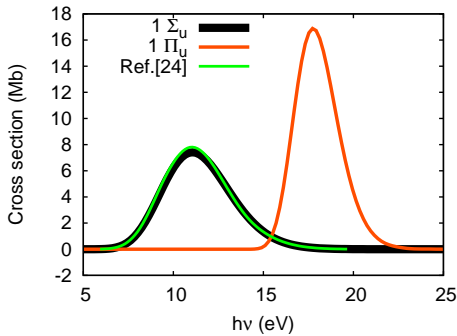}} \\
\end{tabular}
\caption{Photodissociation cross sections for dissociative excitation into the 1s $\Sigma_u$ and 2p $\Pi_u$ electronic states, as calculated presently,
and as calculated for 1s $\Sigma_u$ in Ref.~\cite{dunn}.
\label{disscompare}}
\end{figure}

The prior experimental results~\cite{buschdunn,durup,los,tadj} on photodissociation of H$_2^+$ do not permit a comparison with the present calculation.  
Calculated cross sections for the lowest 1s $\Sigma_u$ photodissociation~\cite{dunn,argyros,saha} and that of the 2p $\Pi_u$~\cite{saha} have been reported.
In Fig.~\ref{disscompare} the cross sections calculated for these final states are shown and that of the $\Sigma_u$ is shown to compare well 
with the result of Dunn calculated near the peak of the cross section within the Born-Oppenheimer approximation~\cite{dunn}.  On this linear scale 
the nonadiabatic contributions are not visible.

\section{Dissociative ionization}

The dissociative ionization flux at a given photon energy may be differentiated with respect to the energy sharing between the electronic
and nuclear degrees of freedom, and with respect to the angular behavior.  
%
The dissociative ionization cross section is obtained by calculating the amplitudes $A_l(k,\kappa)$ for breakup as a function of kinetic energy sharing and electronic angular momentum quantum number,
\begin{equation}
A_l(k,\kappa) = 
\left\langle \Psi_l^-(k,\kappa) \left\vert \mu \right\vert \Psi^0 \right\rangle
\label{ampdef}
\end{equation}

 Presently 
amplitudes $A_l(k,\kappa)$ are calculated exactly, and also using two degrees of approximation, as described below.

\subsection{Exact and approximate amplitude expressions}

  For the fixed-nuclei problem, exact final states were calculated in Ref.~\cite{prolate1_2009} by solving the equation 
\begin{equation}
\Psi^-_l(E) =  \phi_0 + G^+(E) (H-E) \phi_0
\label{exactexpression}
\end{equation}
with the zeroth order wave function a Coulomb wave $\phi_0 = f_l(kr) P_{lm}(\cos \theta)$, $E=\frac{k^2}{2}$,  and the interaction term is
\begin{equation}
(H-E)\phi_0 = \left(\frac{1}{r} - \frac{1}{r_A}-\frac{1}{r_B}\right) \phi_0 \quad .
\label{intterm}
\end{equation}

In contrast, for three body scattering with pairwise interactions Eq.(\ref{exactexpression}) is not valid.
As an alternative to employing an explicit representation of $\Psi^-(E)$, 
stationary phase expressions~\cite{Rudge_TheoryRev,RudgeSeaton_PRPS} that can be implemented in a numerically 
robust way~\cite{accurate,practical,topicalreview04}  have been applied to 
three-body Coulomb breakup problems with two electrons.

However, the prolate spheroidal coordinate system, along with the unequal masses between the electronic
and nuclear degrees of freedom, in general appears to allow Eq.(\ref{exactexpression}) to be implemented such that it yields an accurate final state.  At the
end of the electronic grid in prolate spheroidal coordinates, the electron is always at a greater radius than the nuclei.  Thus, we should not expect to
be able to construct final states for which $\mu_R k < \mu_e \kappa$, for which the protons recoil faster than the electrons and are thereby
shielded from one another by the electron.  Given a maximum of approximately 13.6eV nuclear kinetic energy release, this would indicate
that our results are certainly good above 7.5meV electron energy, a quantity that is not visible on any of the figures.

The final state wave function is thereby calculated as 
\begin{equation}
\Psi^-_l(k,\kappa) = f_l(kr) f_1(\kappa R) + G^+(E) (H-E) f_l(kr) f_1(\kappa R)
\label{psiminus}
\end{equation}
wherein $f_l(kr)$ and $f_1(\kappa R)$ are attractive ($Z=2$) and repulsive ($Z=1$) Coulomb functions in the electronic and nuclear degrees of freedom, energy-normalized.
This is performed in the same manner as in Ref.~\cite{prolate1_2009}.  The final state wave functions so constructed are orthogonal
to the bound rovibronic states for $\mathcal{M}=0$ and 1, to within no more than one part in 10$^{3}$, in general.

With the amplitudes defined as per Eq.(\ref{ampdef}), cross sections differential in the kinetic energy sharing between the electron 
$\frac{k^2}{2}=\epsilon$ and that of the nuclei $\frac{\kappa^2}{2}=E$, at constant total energy $E+\epsilon$, are
%
%
%
%
\begin{equation}
\begin{split}
& \frac{\partial}{\partial E} \sigma(E,\epsilon)\Big\vert_{E+e} =  \frac{4}{3}\pi^2\alpha\omega m \sum_l \left\vert A_l(k,\kappa) \right\vert^2 \\
\end{split}
\label{ampsum}
\end{equation}
wherein $\Psi^-(k,\kappa)$ is energy normalized.

\begin{figure}
\begin{tabular}{c}
\resizebox{0.75\columnwidth}{!}{\includegraphics*[1in,1.3in][5.5in,4.1in]{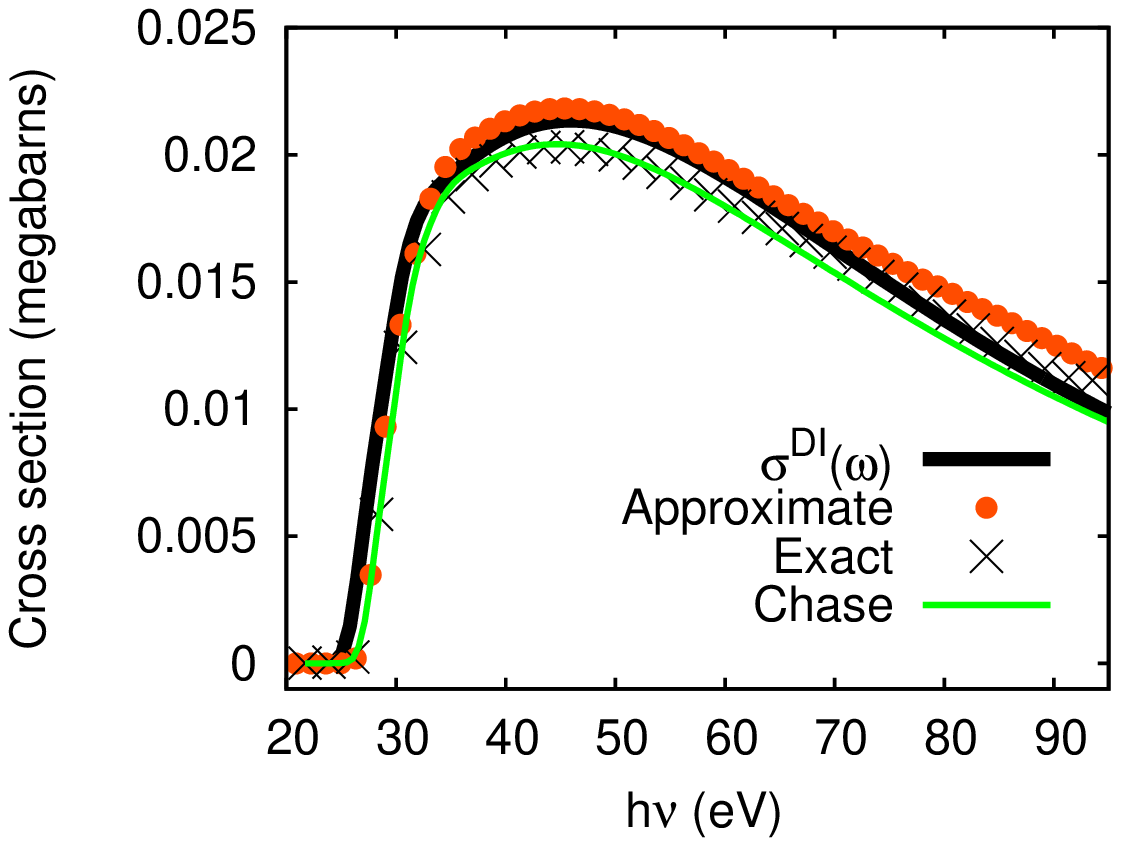}} \\
\resizebox{0.75\columnwidth}{!}{\includegraphics*[0.7in,0.4in][5.5in,4.1in]{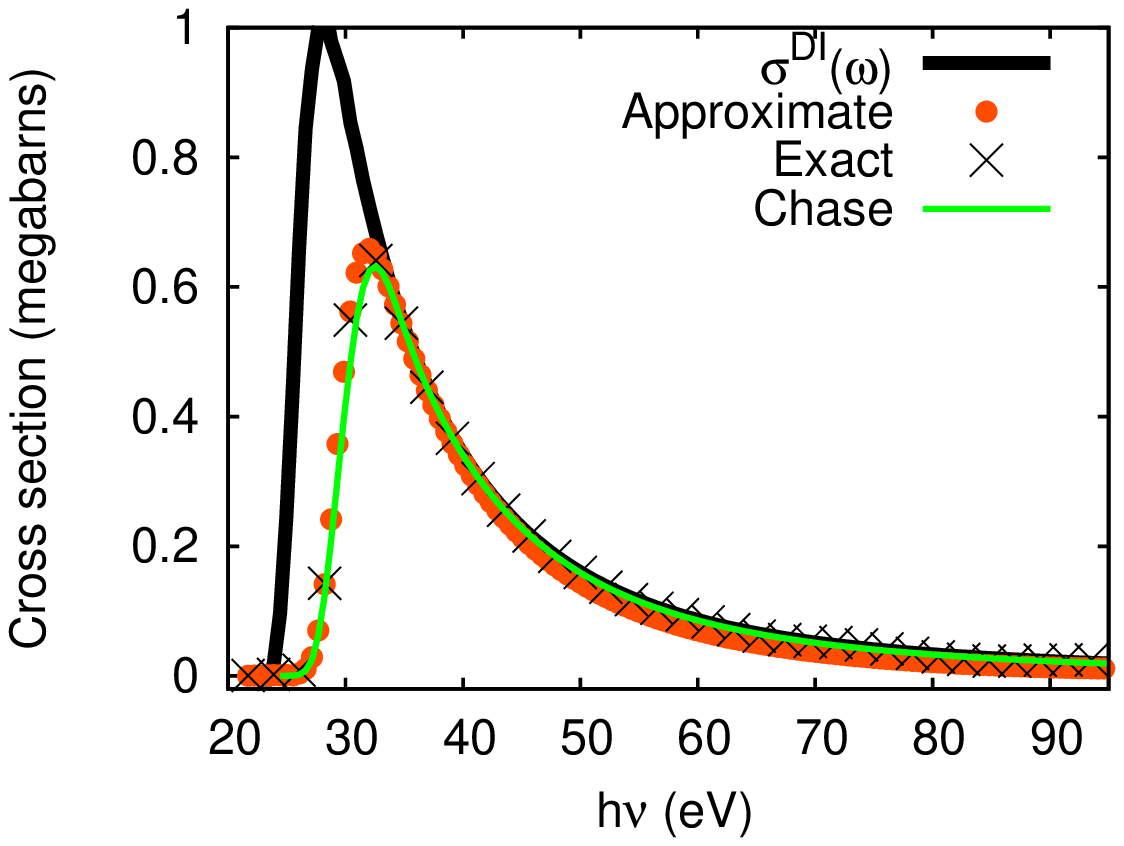}} \\
\end{tabular}
\caption{(Color online) Total cross sections for dissociative ionization calculated in different ways: 
using the flux expression, Eq.~(\ref{spliteq}); using Eq.(\ref{ampdef}) and 
integrating Eq.(\ref{ampsum}) with respect to energy sharing, using the exact states of
Eq.~(\ref{psiminus}), the approximate final states of Eq.(\ref{approxfinal}) or the Chase approximation with Born-Oppenheimer
amplitudes, Eq.~(\ref{chaseexpr}).  
 Top, parallel polarization; bottom, perpendicular.\label{chasex}}
\end{figure}

Approximate final states are often employed in calculations in the literature, and in some contexts simple unperturbed product wave functions $\phi_0$ 
are surprisingly accurate.  For instance, in time-dependent calculations on small atoms and diatomics,
cross sections may be calculated~
\cite{hucolgan, pentuple, h2alignment, hestreaking} by projecting a propagated wave packet onto unperturbed Coulomb wave functions, as long as enough time has elapsed
such that the ionized electrons have escaped beyond the molecule.  In systems containing resonances, this method becomes less tractable the 
longer-lived the resonances are.  A comparison of different amplitude expressions for single and double ionization of two electron systems, 
similar to that presented here for
dissociative ionization of H$_2^+$, can be found in Ref~\cite{comparison}.

Approximate final states $\Psi^-_l(k,\kappa)$ are constructed as products
\begin{equation}
\Psi^-_l(k,\kappa) \approx f_1(\kappa R) \psi^-_l(k^2/2; R)
\label{approxfinal}
\end{equation}
of Coulomb waves in the bond distance and the exact fixed-nuclei scattering state $\psi^-$.  This is therefore a Born-Oppenheimer
representation of the final state.  Finally, the initial state is replaced with its Born-Oppenheimer approximation as well, such that the amplitudes are
the matrix elements of the Born-Oppenheimer amplitudes with respect to the initial and final vibrational states, which expression is 
called the Chase approximation~\cite{chase}:
\begin{equation}
A_l(k,\kappa) \approx \int dR f_1(\kappa R) \chi_0(R) A_l(k; R) \quad ,
\label{chaseexpr}
\end{equation}
in which $\chi_0$ is the ground vibrational Born Oppenheimer state.


\begin{figure}
\begin{tabular}{c}
\resizebox{0.75\columnwidth}{!}{\includegraphics*[1in,0.5in][5.6in,4.1in]{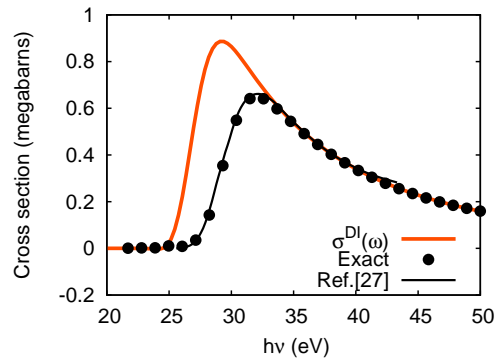}} \\
\end{tabular}
\caption{(Color online) Dissociative ionization cross section calculated exactly, using Eq.(\ref{spliteq}), and as in Ref.~\cite{poots}. \label{ioncompare}}
\end{figure}

\begin{figure*}
\begin{center}
\begin{tabular}{cc}
\resizebox{0.7\columnwidth}{!}{\includegraphics*[0.8in,1.2in][5.7in,4.2in]{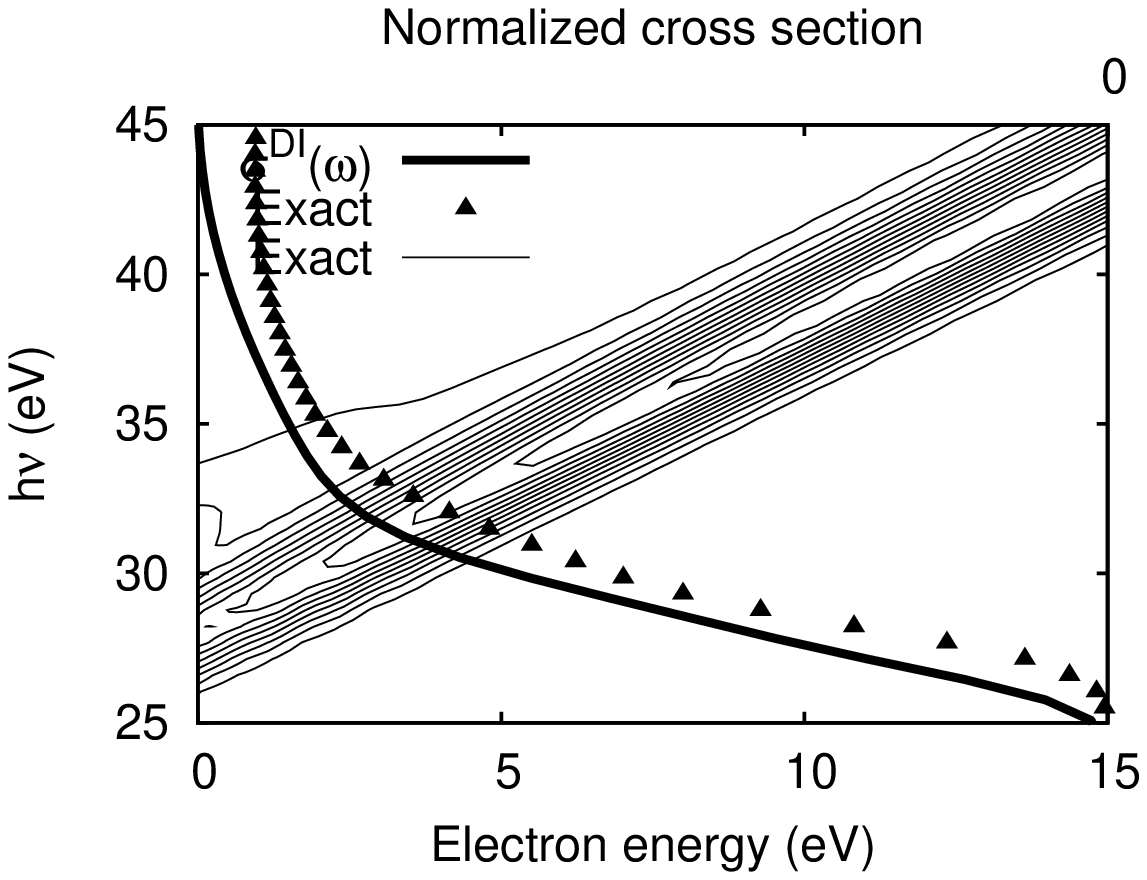}} &
\resizebox{0.7\columnwidth}{!}{\includegraphics*[0.8in,1.2in][5.7in,4.2in]{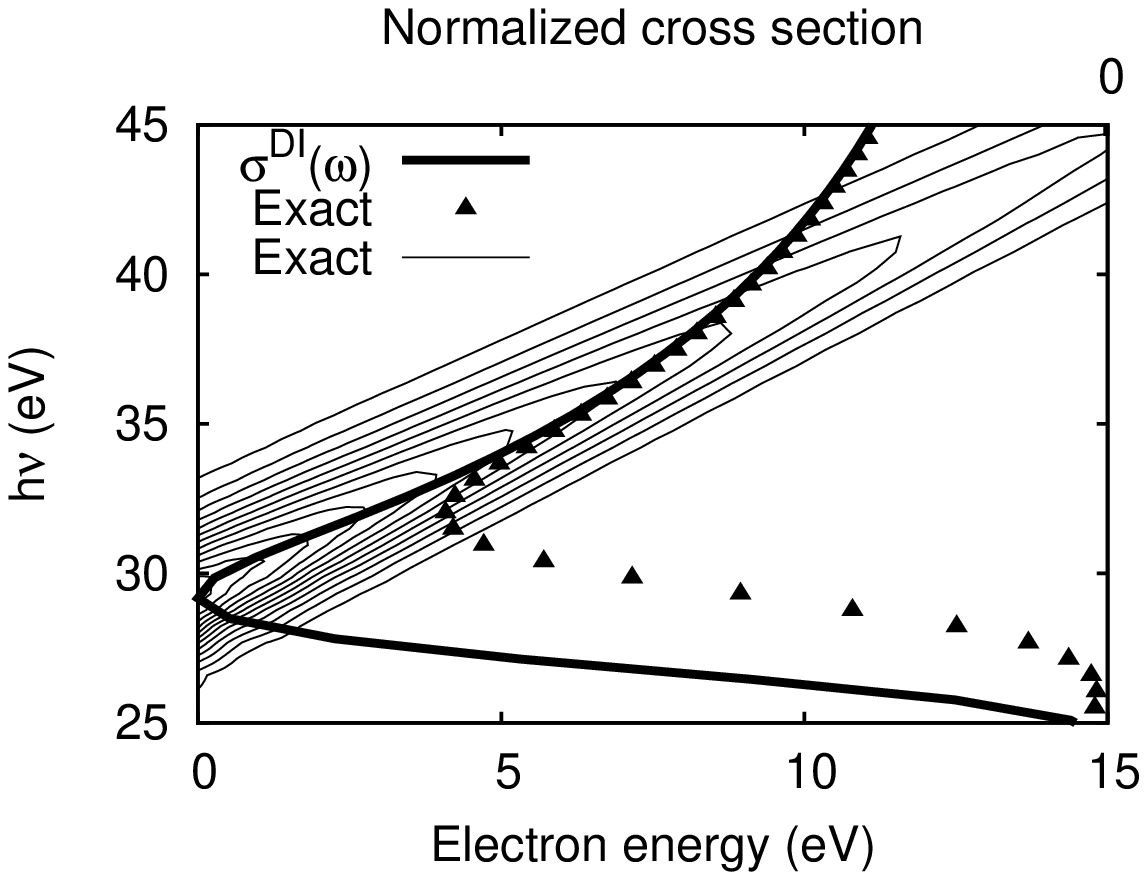}} \\
\resizebox{0.7\columnwidth}{!}{\includegraphics*[0.8in,1.2in][5.7in,3.8in]{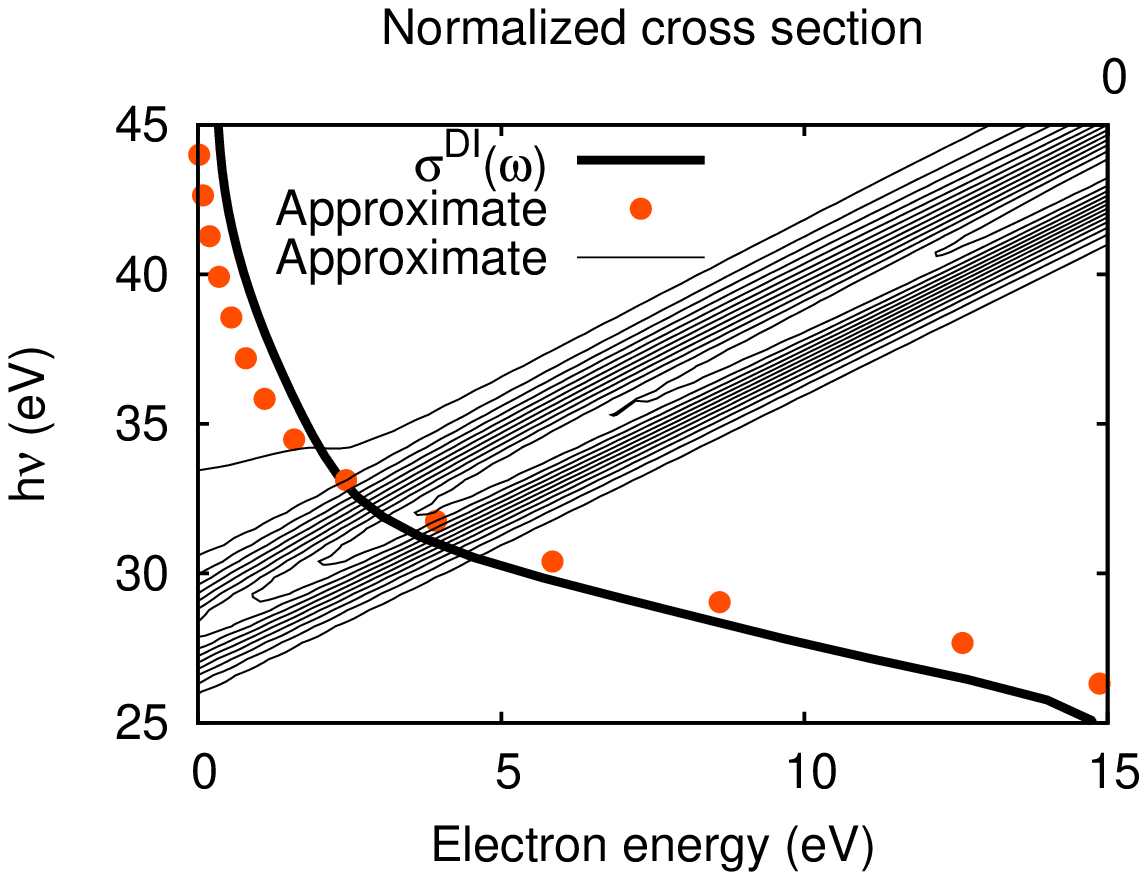}} &
\resizebox{0.7\columnwidth}{!}{\includegraphics*[0.8in,1.2in][5.7in,3.8in]{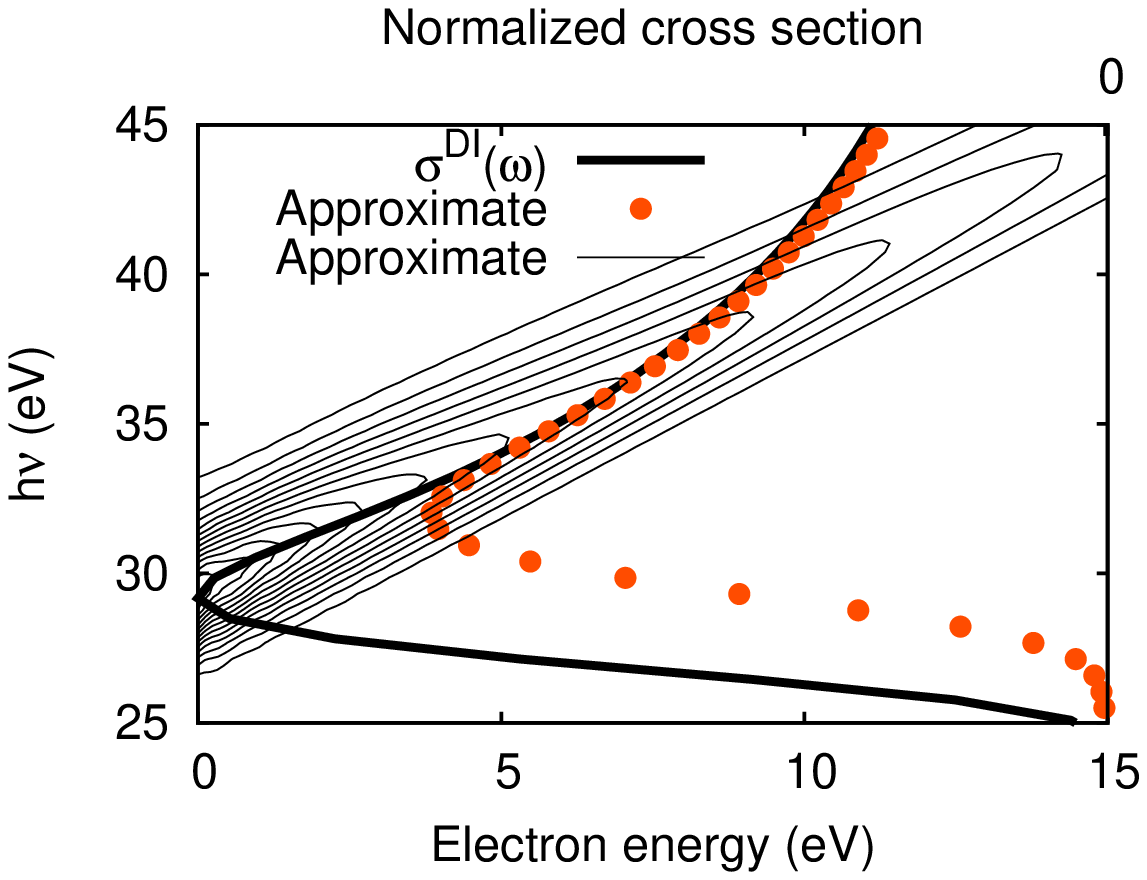}} \\
\resizebox{0.7\columnwidth}{!}{\includegraphics*[0.8in,0.5in][5.7in,3.8in]{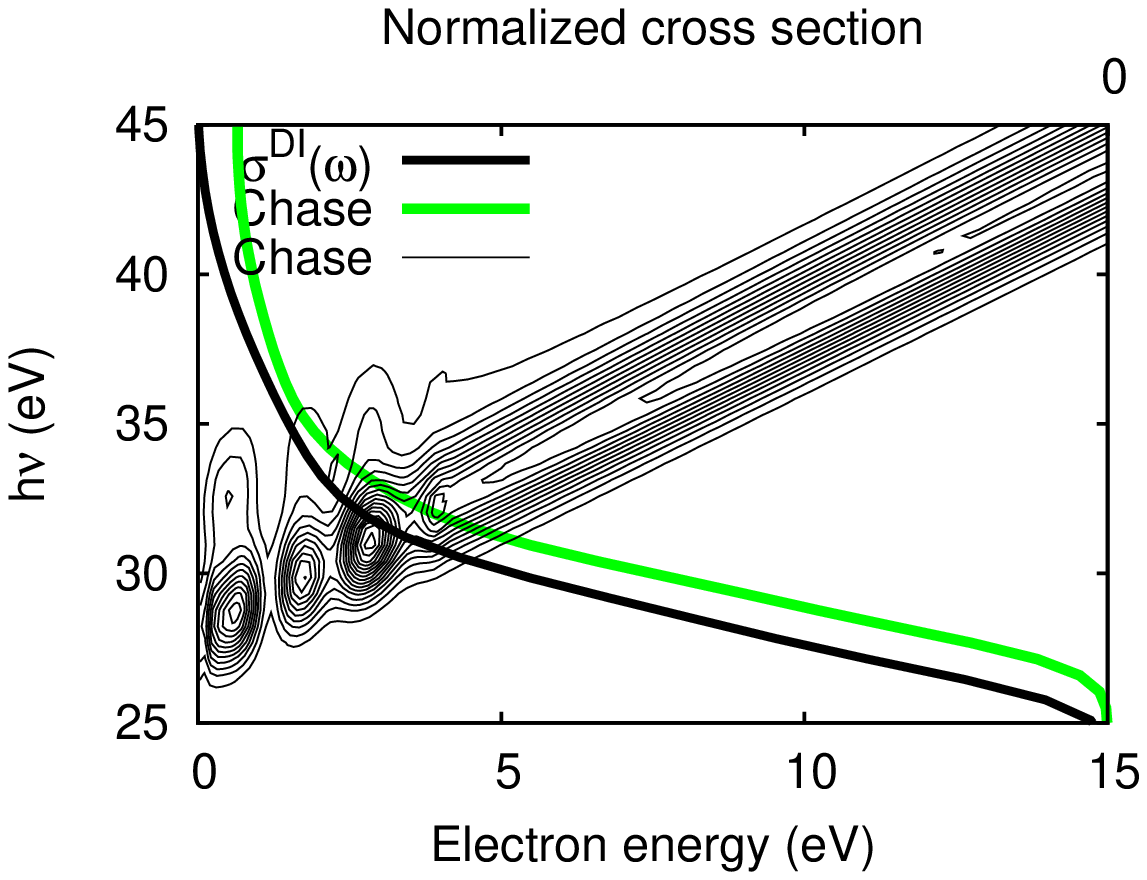}} &
\resizebox{0.7\columnwidth}{!}{\includegraphics*[0.8in,0.5in][5.7in,3.8in]{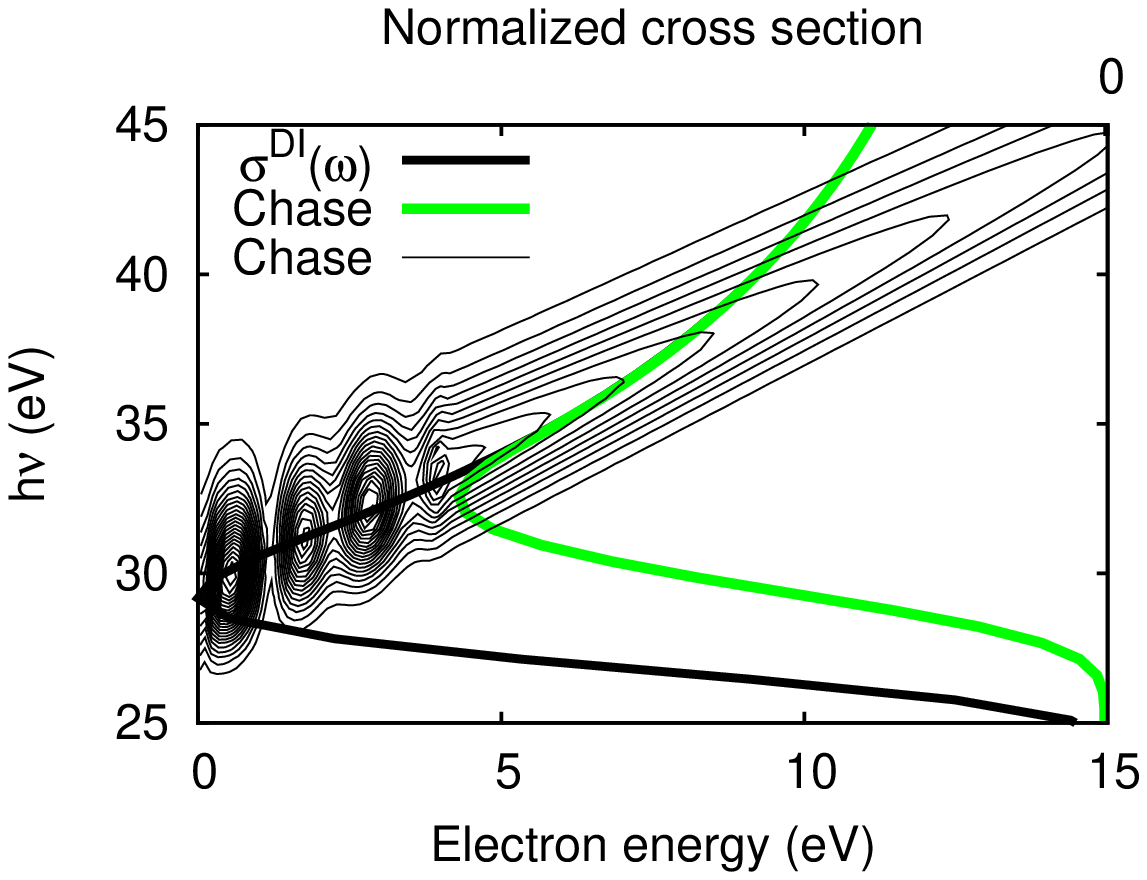}} \\
\end{tabular}
\end{center}
\caption{(Color online) Distributions of kinetic energy between the electron and nuclei
 calculated exactly (top), using Born-Oppenhemer final states (middle) and using Born-Oppenheimer
initial and final states (bottom, Chase approximation).
The cross section differential in energy sharing, Eq.(\ref{ampsum}), is plotted with contours
as a function of electron energy and photon energy.  The total cross section
without regard to energy sharing is plotted vertically, as a function of photon energy.  The
result of integrating the differential cross section is plotted with different styles as in Fig.~\ref{chasex}, and 
the dissociative ionization cross section as calculated via Eq.(\ref{spliteq}) is plotted bold black.
\label{kedist}}
\end{figure*}

\subsection{Results}

Cross sections for dissociative ionization are shown in Fig.~\ref{chasex}.  For these total cross sections, there is very little difference 
among the various results calculated using the different amplitude expressions, exact and approximate.  However, in perpendicular
polarization, there is a large
discrepancy between these results and $\sigma^{DI}(\omega)$ as defined by Eq.\ref{spliteq}.  The origin of this discrepancy is unclear,
but it calls into question the division of the cross section as defined by that equation.  Further study of this discrepancy is therefore indicated.
The results calculated via the amplitude expressions should be regarded as reliable, due to the fact that one of them has been calculated in a formally
exact manner.

The dissociative ionization cross section is compared to the calculation of Ref.~\cite{poots}, the exact analytic fixed-nuclei result convolved over
the initial vibrational wave function, in Fig.~\ref{ioncompare}.  This and the present calculation are also in agreement with the prior Born-Oppenheimer
results~\cite{undulations,billtom,larkins,yabu}. The cross section is overwhelmingly dominated by the perpendicular component, and the perpendicular component of the total 
dissociative ionization cross section is affected little by inclusion of the internuclear coordinate.

The distributions of kinetic energy between the electrons and nuclei are shown in Fig.~\ref{kedist}.  In these figures the exact, approximate, 
and Born-Oppenheimer (Chase approximation) 
results are compared.  One can see that for electron energies above 5eV, the three results are substantially in agreement.  Below 5eV, however, the
Chase approximation yields qualitatively incorrect behavior, yielding strong minima in the cross section as a function of energy sharing whereas in the exact result there are none. 
In terms of the distribution of kinetic energy between the electron and the nuclei, 
the results for approximate Born-Oppenheimer final states do not significantly differ from the exact ones.

\begin{figure*}
\begin{tabular}{ll}
\rotatebox{90}{TDCS}
&
\begin{tabular}{cccc}
\resizebox{0.25\textwidth}{!}{\includegraphics*[1.5in,1.1in][5.6in,3.9in]{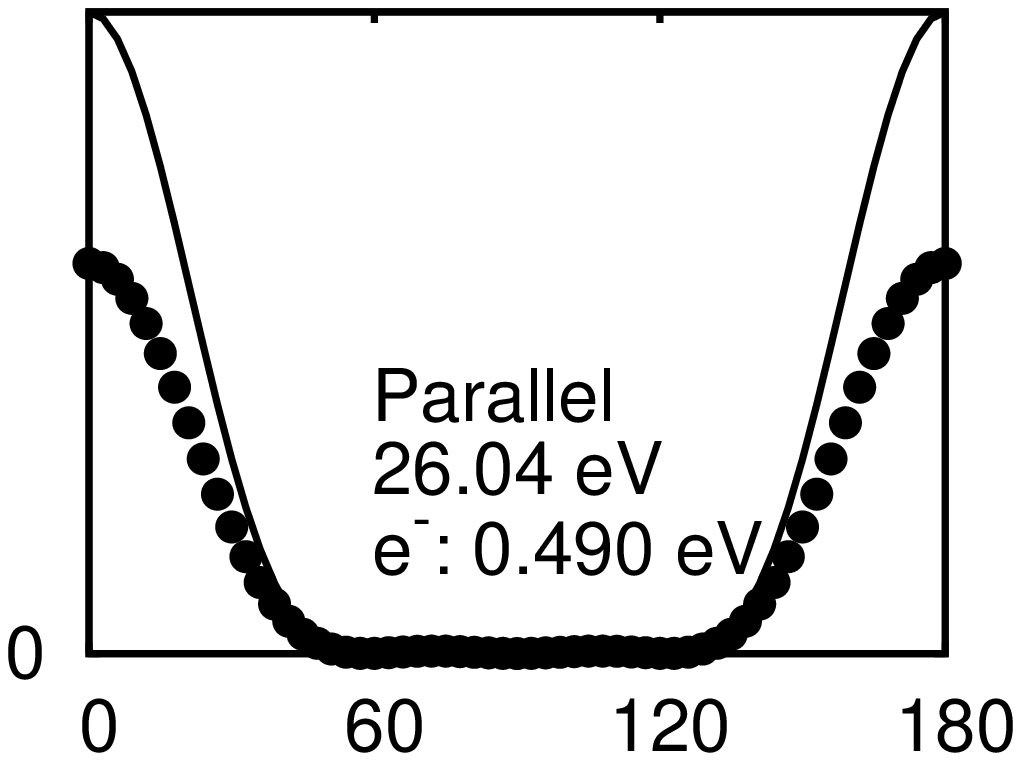}} &
\resizebox{0.24\textwidth}{!}{\includegraphics*[1.7in,1.1in][5.6in,3.9in]{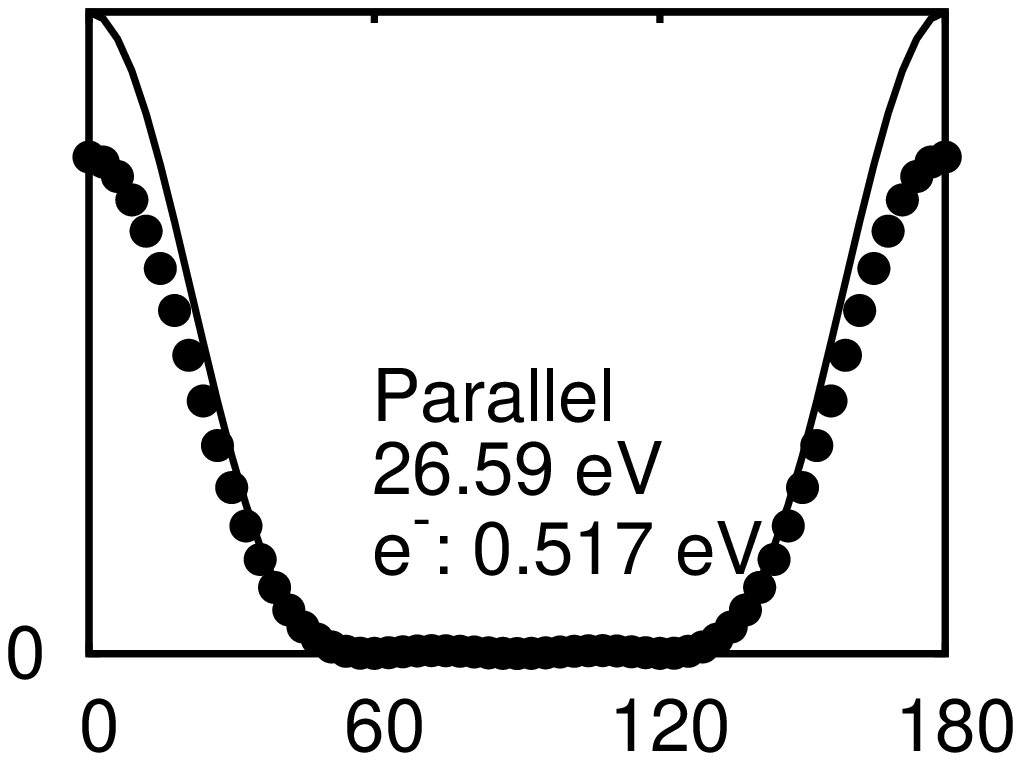}} &
\resizebox{0.24\textwidth}{!}{\includegraphics*[1.7in,1.1in][5.6in,3.9in]{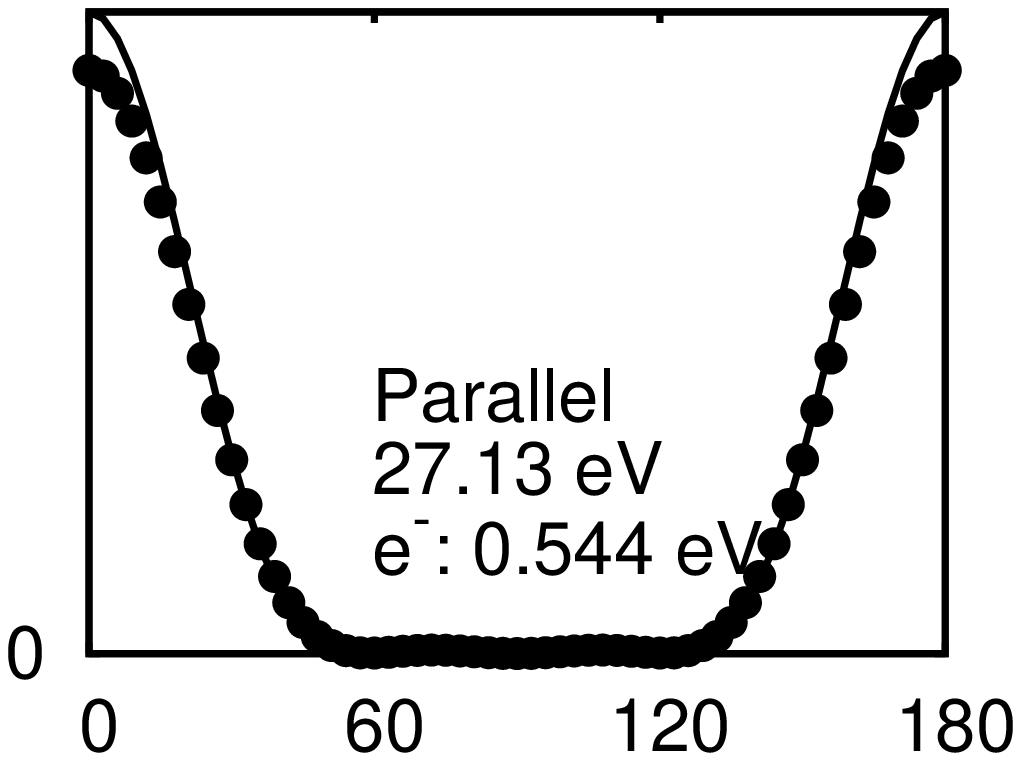}} &
\resizebox{0.24\textwidth}{!}{\includegraphics*[1.7in,1.1in][5.6in,3.9in]{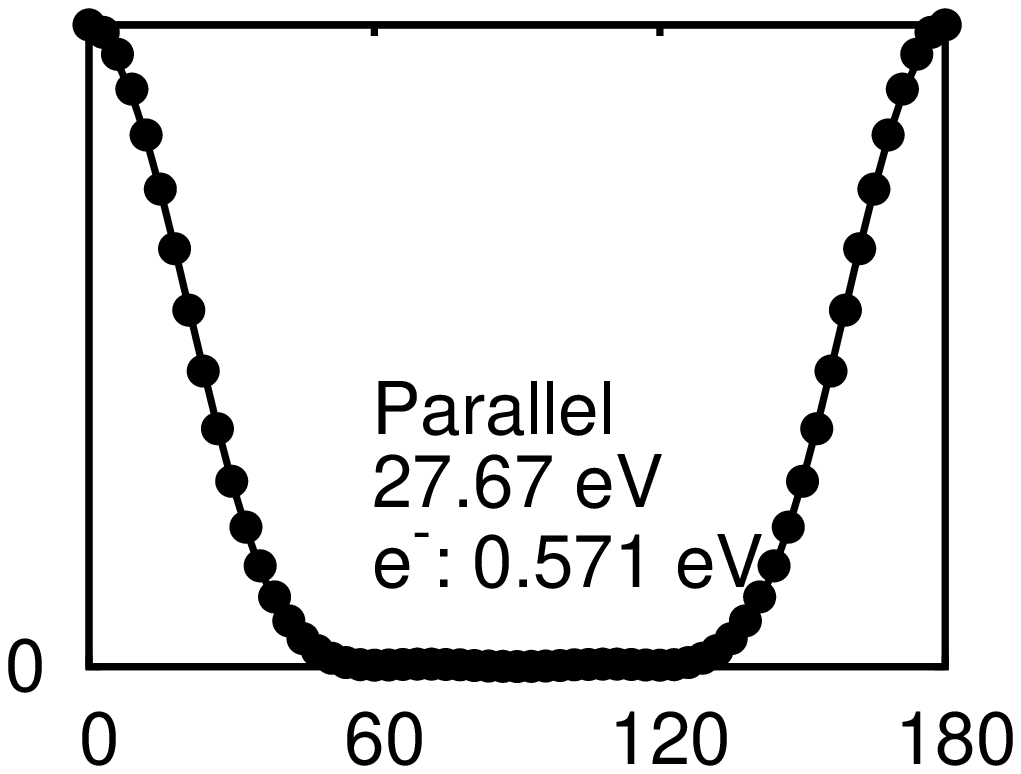}} \\
\resizebox{0.25\textwidth}{!}{\includegraphics*[1.5in,1.1in][5.6in,3.9in]{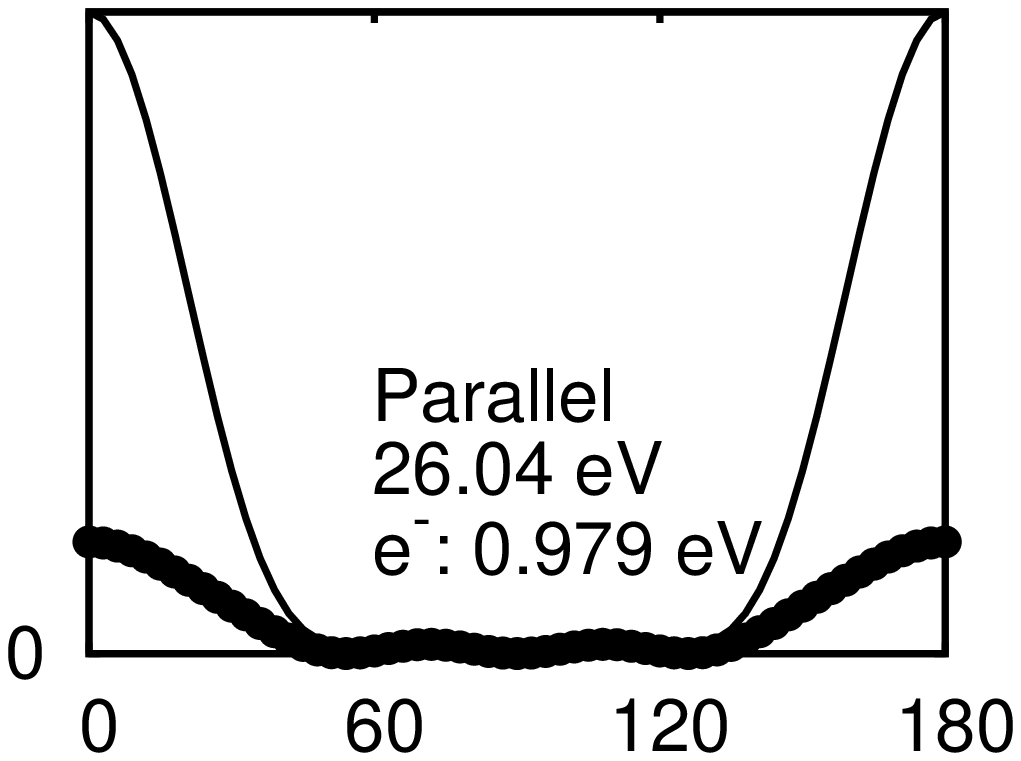}} &
\resizebox{0.24\textwidth}{!}{\includegraphics*[1.7in,1.1in][5.6in,3.9in]{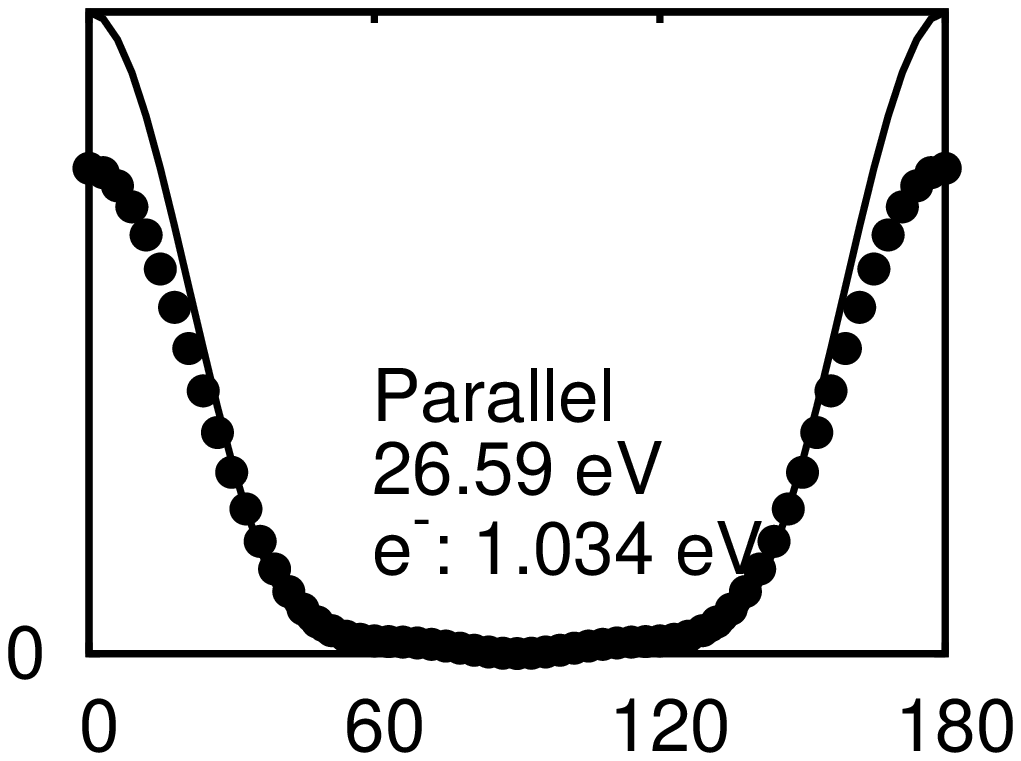}} &
\resizebox{0.24\textwidth}{!}{\includegraphics*[1.7in,1.1in][5.6in,3.9in]{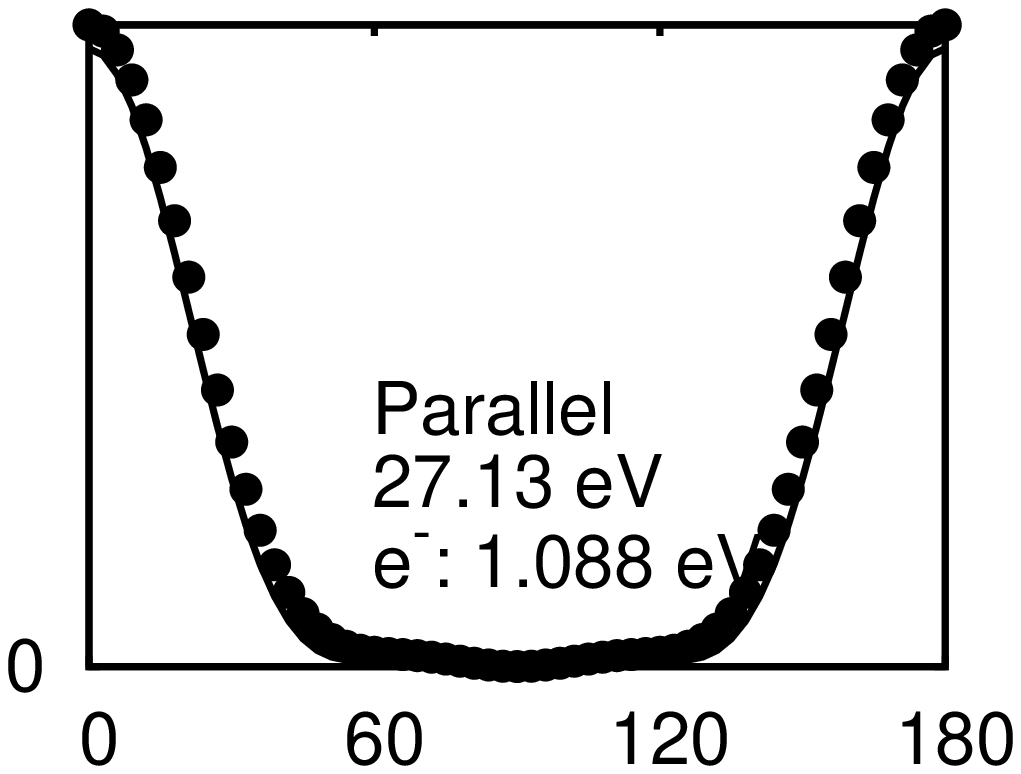}} &
\resizebox{0.24\textwidth}{!}{\includegraphics*[1.7in,1.1in][5.6in,3.9in]{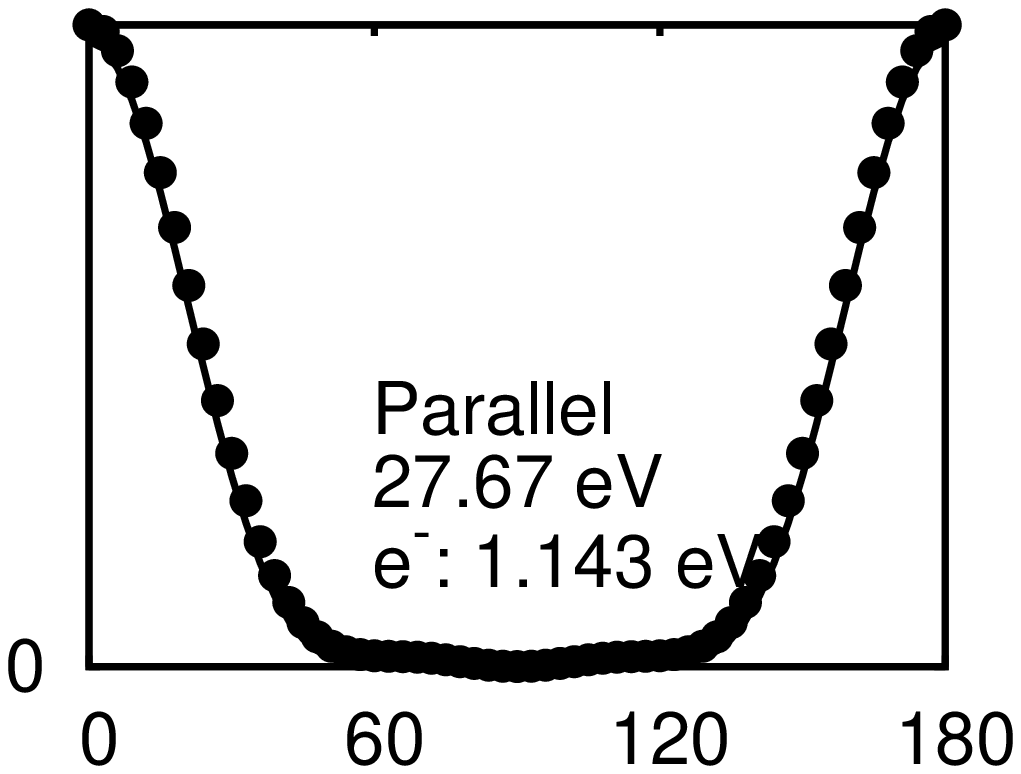}} \\
\resizebox{0.25\textwidth}{!}{\includegraphics*[1.5in,1.1in][5.6in,3.9in]{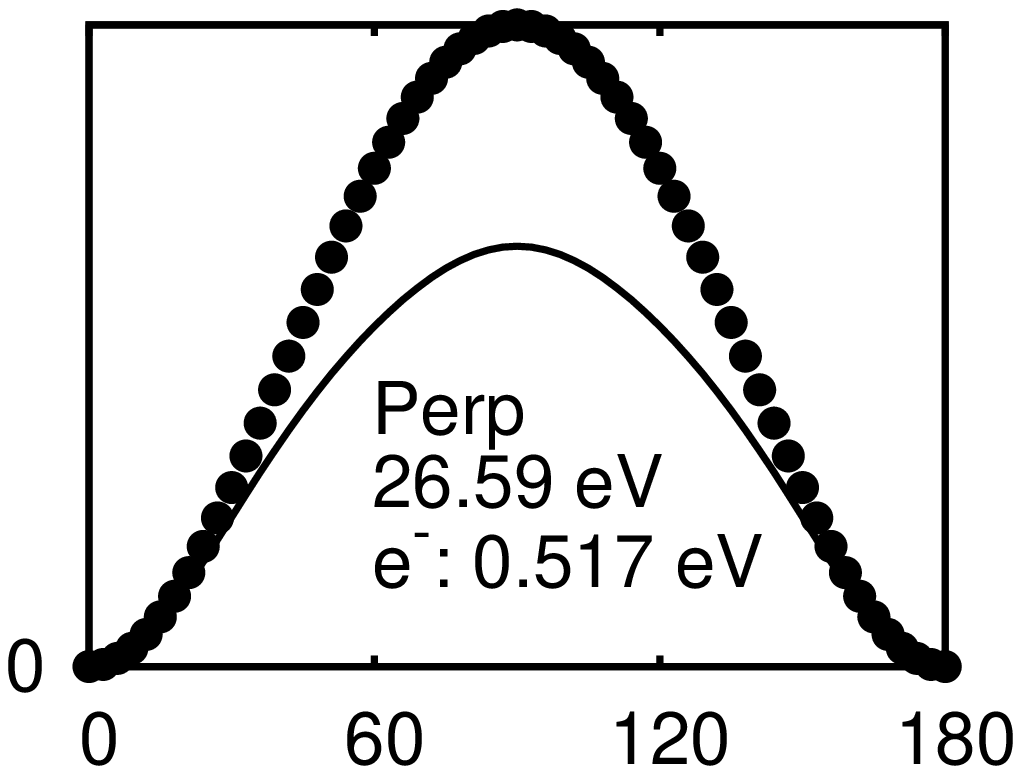}} &
\resizebox{0.24\textwidth}{!}{\includegraphics*[1.7in,1.1in][5.6in,3.9in]{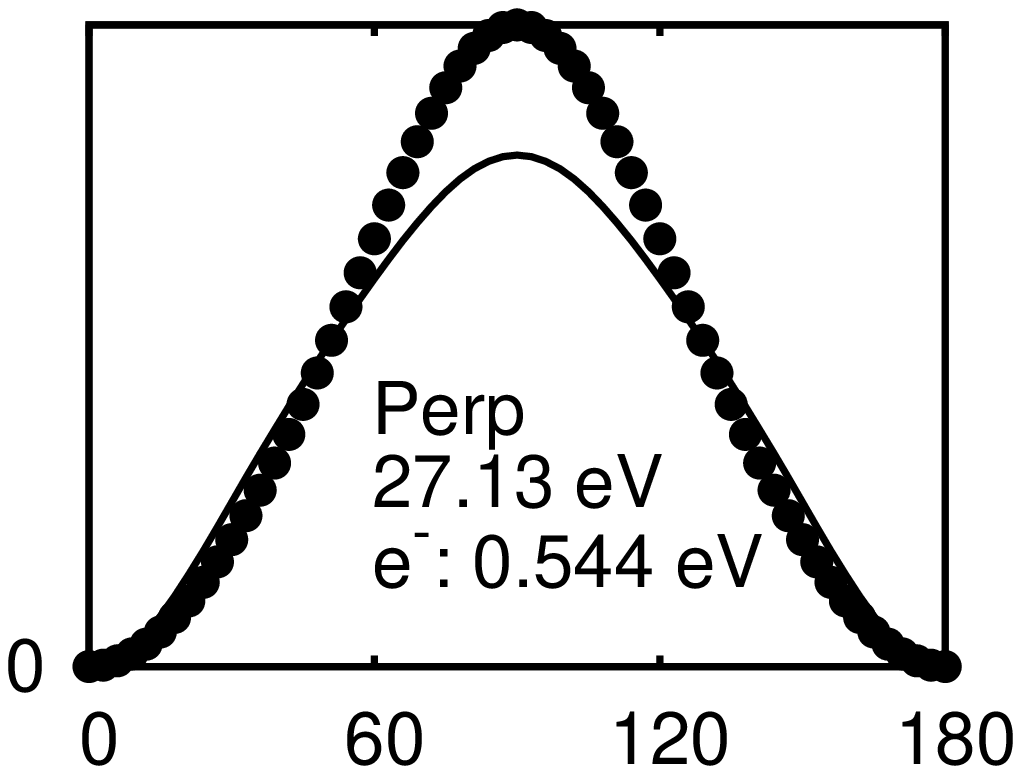}} &
\resizebox{0.24\textwidth}{!}{\includegraphics*[1.7in,1.1in][5.6in,3.9in]{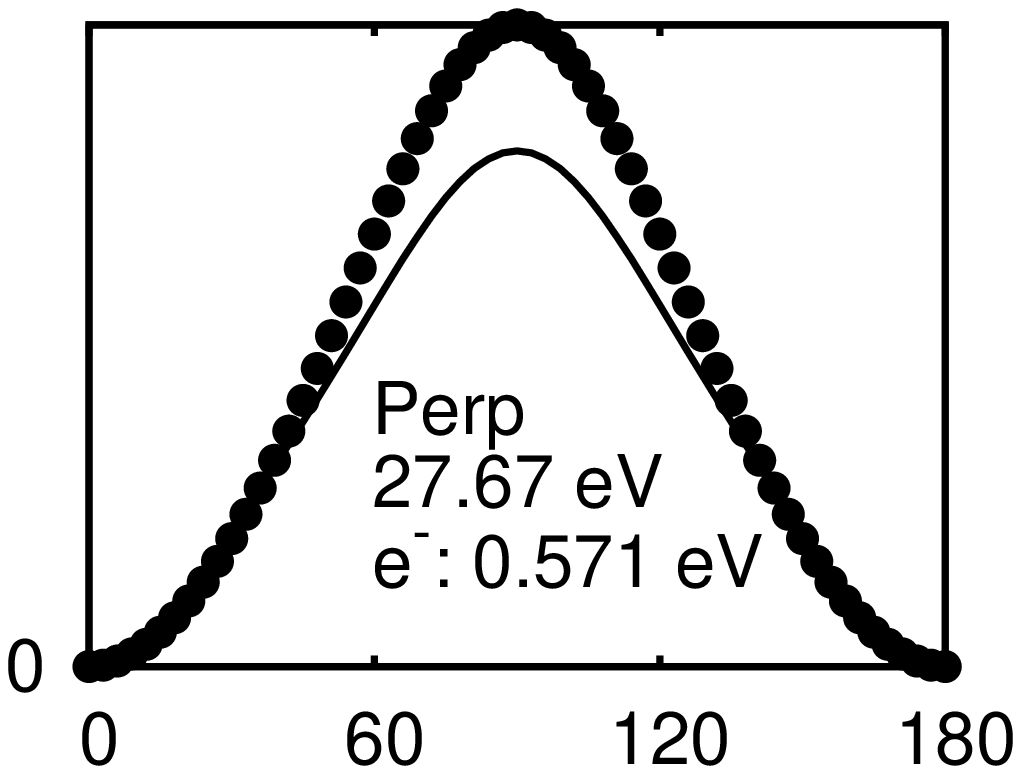}} &
\resizebox{0.24\textwidth}{!}{\includegraphics*[1.7in,1.1in][5.6in,3.9in]{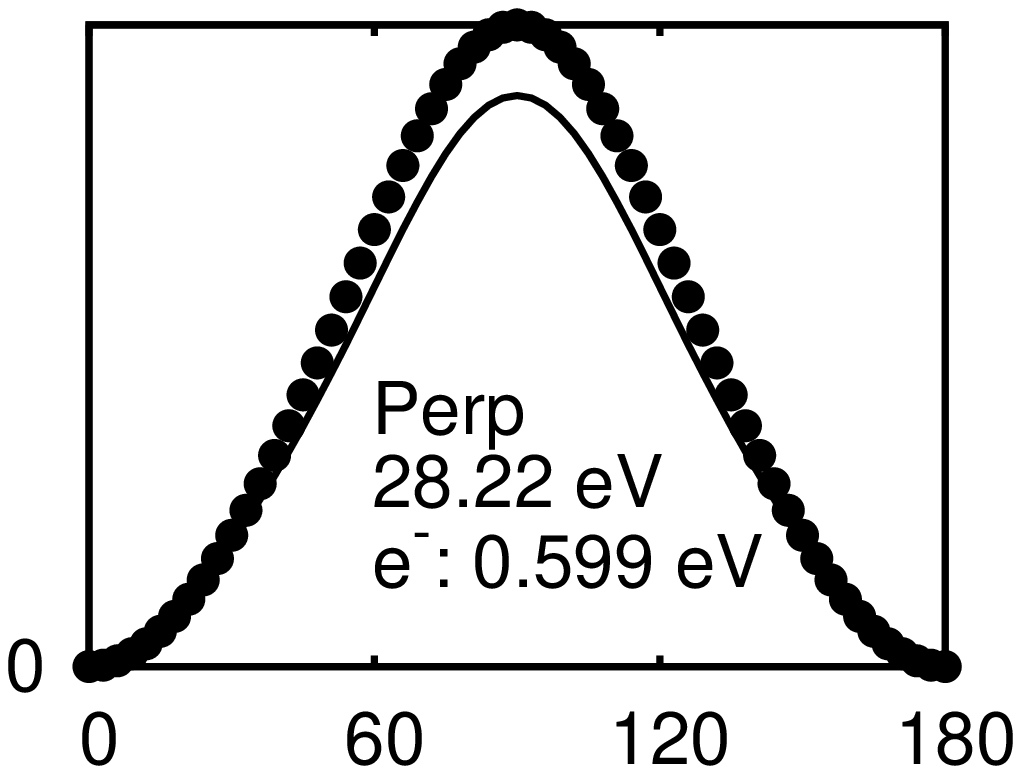}} \\
\resizebox{0.25\textwidth}{!}{\includegraphics*[1.5in,0.5in][5.6in,3.9in]{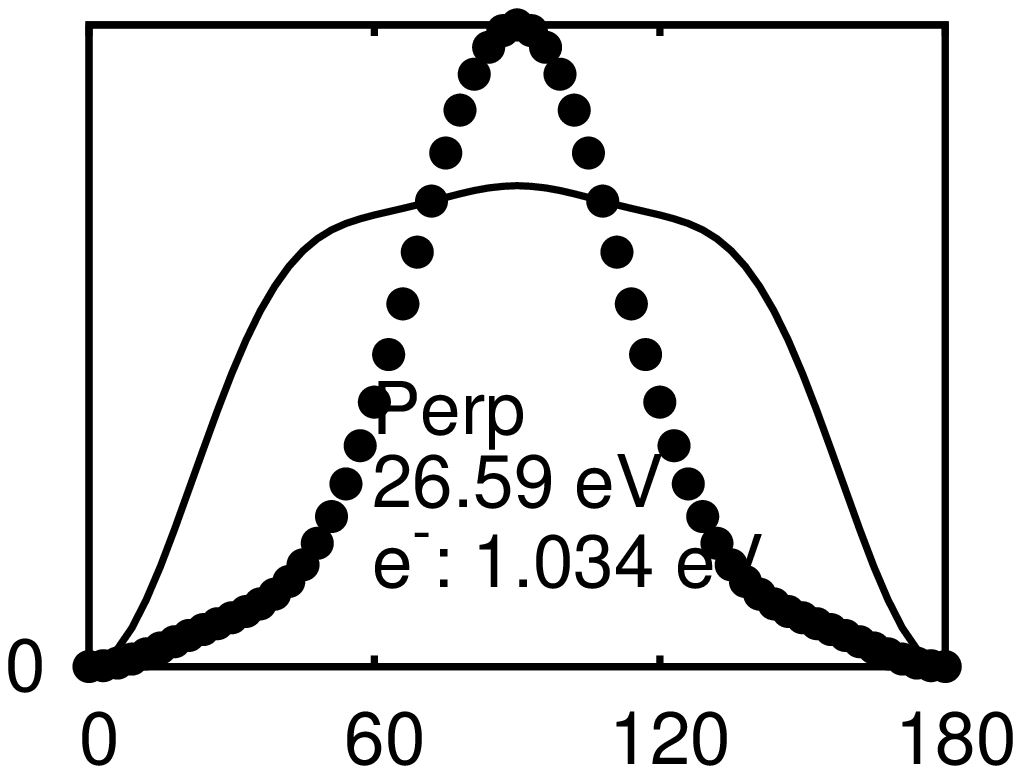}} &
\resizebox{0.24\textwidth}{!}{\includegraphics*[1.7in,0.5in][5.6in,3.9in]{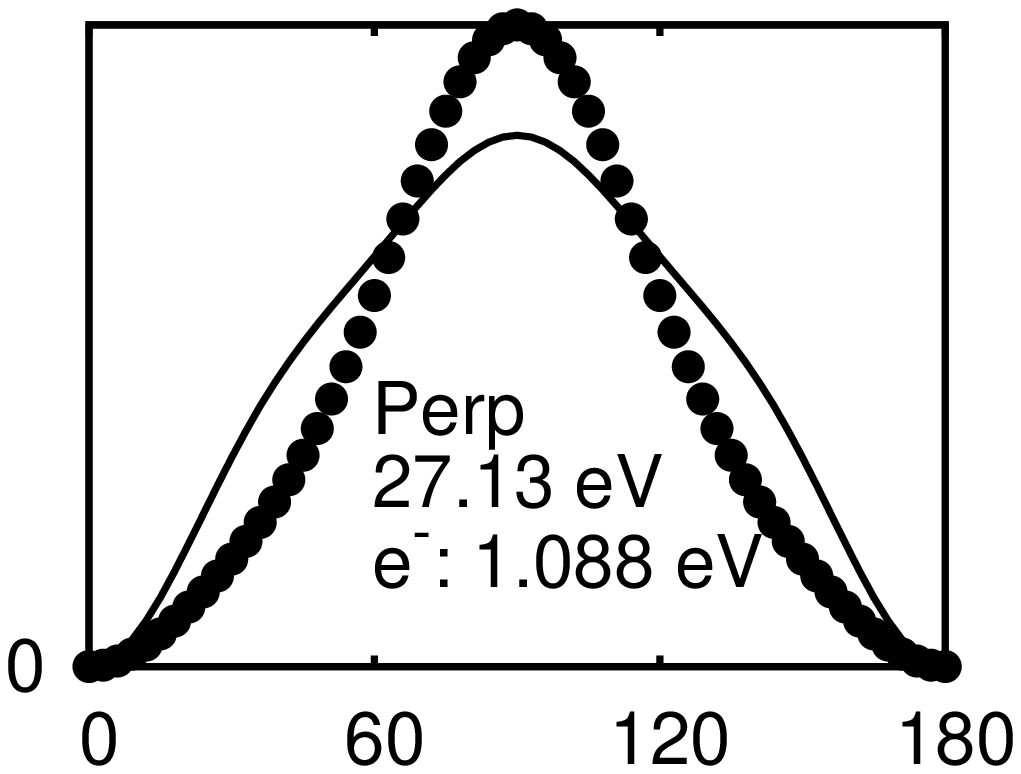}} &
\resizebox{0.24\textwidth}{!}{\includegraphics*[1.7in,0.5in][5.6in,3.9in]{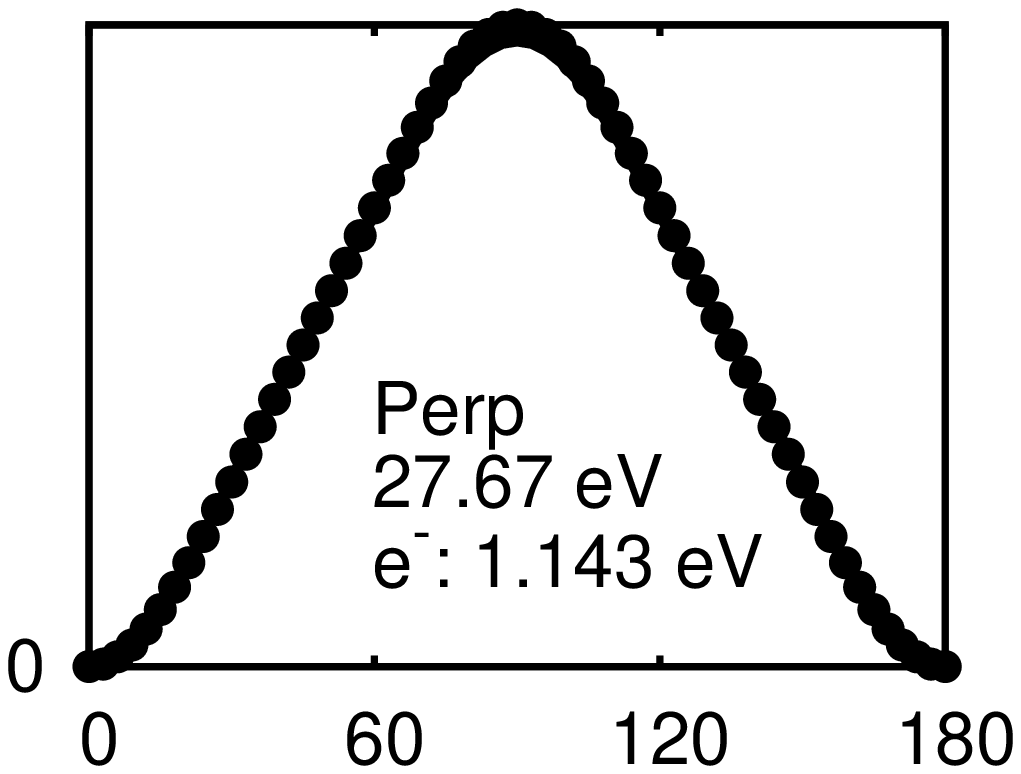}} &
\resizebox{0.24\textwidth}{!}{\includegraphics*[1.7in,0.5in][5.6in,3.9in]{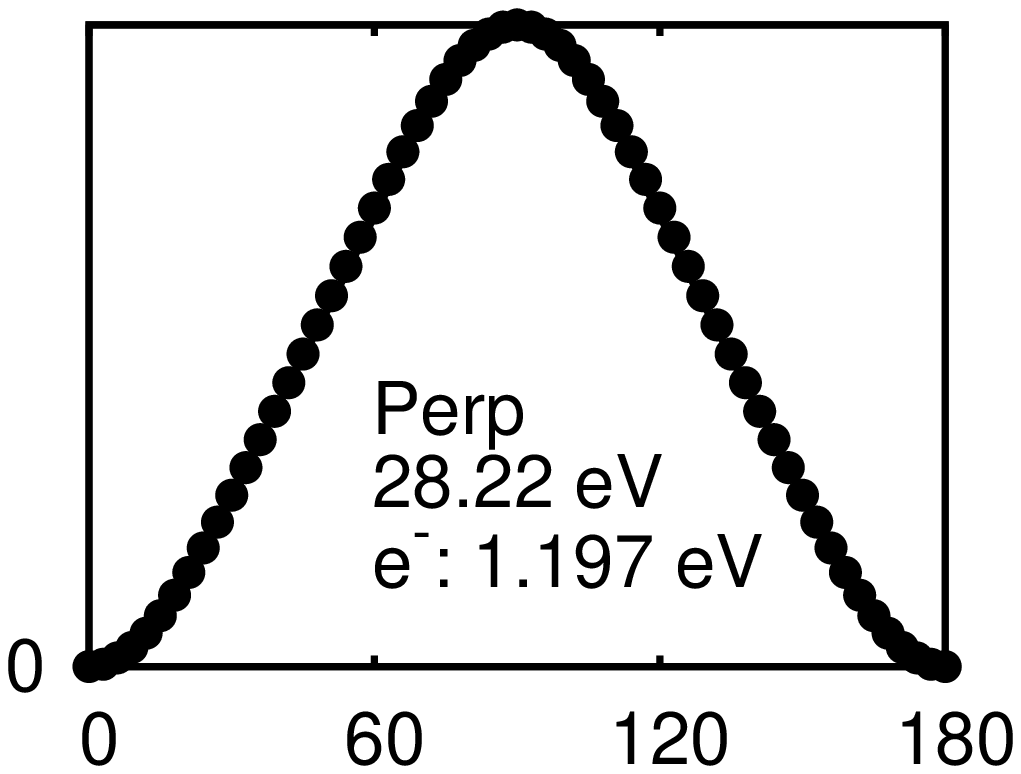}} \\
\multicolumn{4}{c}{Angle (degrees)}
\end{tabular}
\end{tabular}
\caption{Triply differential cross sections for dissociative photoionization calculated exactly (dots) and with approximate final states (lines),
with each panel plotted on an arbitrary scale, the two results in each panel internormalized.  On each panel the photon energy and the outgoing
electron energy are indicated.  \label{tdcs}}
\end{figure*}

However, when the full, triply differential cross section is calculated, there are clear differences, indicating that an exact nonadiabatic treatment is 
indeed necessary to fully describe the breakup of H$_2^+$.  In Fig.~\ref{tdcs} the triply differential cross section, differential with respect to energy,
energy sharing, and the relative angle of ionization and dissociation is plotted near onset and for low electron energies.  In general, in parallel polarization,
these figures show that the approximate treatment with Born-Oppenheimer final states somewhat overestimates these low electron kinetic energy cross sections.  
However, the shape of the TDCS for parallel polarization -- that is to say,
the relative magnitude and phases of the partial waves contributing to it -- is in agreement and nearly constant over all energies for both treatments.  
In contrast, for perpendicular polarization
there are substantial differences in the shape of the TDCS obtained via the exact and approximate final state treatments.  This indicates that nonadiabatic effects
are important for a completely accurate description of the dynamics.

\section{Conclusion}

Full nonadiabatic calculations of the cross sections for breakup of the H$_2^+$ cation by photon impact have been presented.  In the case of dissociative
ionization the exact result has been critically compared to approximate ones, and it was shown that the Born-Oppenheimer approximation gives cross sections
differential in energy sharing that are very close to the exact result.  However, an accurate calculation of the fully differential cross section requires the
full nonadiabatic treatment.   The use of the described flux formalism to calculate the dissociative ionization the cross section,
Eq.(\ref{spliteq}), is called into question due to its disagreement with the formally exact result in perpendicular polarization.  
 In the case of dissociative excitation, the cross sections have been calculated over six orders of magnitude, revealing the influence of nonadiabatic coupling. 

\section{Acknowledgements}
This work  was supported by the US Department of Energy Office of Basic Energy Sciences, Division of Chemical Sciences Contract  DE-AC02-05CH11231.

\bibliography{h2plus}

\end{document}